\documentclass{emulateapj-rtx4}
\usepackage{amssymb, amsmath, epsfig} 

\newcommand{\grad}{{\mathbf{\nabla}}}
\newcommand{\TS}{{T_{\rm S}}}
\newcommand{\TCMB}{{T_{\rm CMB}}}
\newcommand{\Tk}{{T_{\rm K}}}
\newcommand{\calM}{{\cal M}}
\newcommand{\calbfM}{{\boldsymbol{\cal M}}}
\newcommand{\bfk}{{\boldsymbol{k}}}

\newcommand{\bfv}{{\boldsymbol{v}}}

\newcommand{\bc}{{\rm bc}}
\newcommand{\bfr}{{\boldsymbol r}}

\newcommand{\cm}{{\rm cm}}
\newcommand{\bfx}{{\boldsymbol x}}
\newcommand{\nhat}{{\hat{\boldsymbol{n}}}}

\newcommand{\cMpc}{{\rm Mpc}}

\newcommand{\MHz}{{\rm MHz}}

\newcommand{\rhoSFR}{{\dot \rho_{\rm SFR}}}

\newcommand{\Msun}{{M_{\odot}}}

\def\apjl{ApJL}
\def\apjs{ApJS}

\def\aap{{A\&A}}

\begin{document}

\title{The impact of the supersonic baryon--dark matter velocity difference on the $z\sim 20$ 21cm background}
\author{Matthew McQuinn\altaffilmark{1}$^,$\altaffilmark{2} \& Ryan M.\ O'Leary\altaffilmark{1}$^,$\altaffilmark{2} }

\altaffiltext{1} {Einstein Fellows\\}
\altaffiltext{2} {Department of Astronomy, University of California, Berkeley, CA 94720, USA\\}

\begin{abstract}
 Recently, Tseliakhovich and Hirata (2010) showed that during the cosmic Dark Ages the baryons were typically moving supersonically with respect to the dark matter with a spatially variable Mach number.  Such supersonic motion may source shocks that inhomogeneously heat the Universe.  This motion may also suppress star formation in the first halos.  Even a small amount of coupling of the 21cm signal to this motion has the potential to vastly enhance the 21cm brightness temperature fluctuations at $15 \lesssim z \lesssim 40$ as well as to imprint distinctive acoustic oscillations in this signal.  We present estimates for the size of this coupling, which we  calibrate with a suite of cosmological simulations of the high-redshift Universe using the GADGET and Enzo codes.  Our simulations, discussed in detail in a companion paper, are initialized to self-consistently account for gas pressure and the dark matter--baryon relative velocity, $v_{\rm bc}$ (in contrast to prior simulations).  We find that the supersonic velocity difference dramatically suppresses structure formation on $10-100~$comoving~kpc scales, it sources shocks throughout the Universe, and it impacts the accretion of gas onto the first star-forming minihalos (even for halo masses as large as $10^7~\Msun$).  However, prior to reheating by astrophysical sources, we find that the $v_{\rm bc}$--sourced temperature fluctuations can contribute only as much as $\approx 10\%$ of the fluctuations in the 21cm signal.  We do find that $v_{\rm bc}$ in certain scenarios could source an ${\cal O}(1)$ component in the power spectrum of the 21cm background on observable scales via the X-ray (but not ultraviolet) backgrounds produced once the first stars formed.  In a scenario in which $\sim 10^6~\Msun$ minihalos reheated the Universe via their X-ray backgrounds, we find that the pre-reionization 21cm signal would be larger than previously anticipated and exhibit more significant acoustic features.  Such features would be a direct probe of the first stars and black holes.  In addition, we show that structure formation shocks are unable to heat the Universe sufficiently to erase a strong 21cm absorption trough at $z\sim 20$ that is found in most models of the sky--averaged 21cm intensity.  
\end{abstract}


\section{Introduction}

Sometime between the redshifts of $30$ and $10$, it is thought that the Universe transitioned from a pristine landscape, where the distribution of matter was calculable from the cosmological initial conditions alone, to a vastly more complex system in which stars abound.  Once these stars formed, within just a few hundred million years  they had reheated and reionized the entire Universe.  These cosmic times can be observed directly via redshifted 21cm radiation from neutral hydrogen.   In fact, predictions are that the 21cm line is visible in absorption against the Cosmic Microwave Background (CMB) from $15\lesssim z \lesssim 40$, with a larger amplitude for its mean brightness temperature than from any other cosmological epoch (e.g., \citealt{pritchard08, furlanettoohbriggs}).  Motivated by these predictions, the radio instruments LEDA, DARE, and LOFAR are being designed to observe 21cm emission from the era prior to reionization, the so-called cosmic Dark Ages \citep{burns11, bernardi12, harker10}.\footnote{\url{www.cfa.harvard.edu/LEDA}; \url{http://lunar.colorado.edu/dare/}; \url{http://www.lofar.org/}}  These efforts are in addition to those aiming to detect the 21cm signal from the Reionization Epoch (targeting $z\sim 6-12$), which includes EDGES, PAPER, MWA, LOFAR (with their high band antennae), and GMRT \citep{bowman10, parsons10, harker10, paciga11}.  
EDGES, LEDA, and DARE are targeting the sky-averaged 21cm signal, whereas the other instruments are attempting to measure fluctuations in the 21cm intensity on $\sim10~$comoving~Mpc scales.  Plans for the next generation of instruments are also in development.\footnote{\url{https://www.cfa.harvard.edu/$\sim$lincoln/astro2010.hera.pdf};\\  \url{http://www.skatelescope.org/}}

Yet, despite the prospects for observations of 21cm emission from $10\lesssim z \lesssim 40$, key theoretical questions regarding these times remain unanswered -- even questions which are independent of the large uncertainties inherent to modeling the astrophysical processes that impact the 21cm signal.   For example, it is unclear whether weak structure formation shocks would have significantly heated the cosmic gas \citep{gnedin04, furlanettoshocks}.  
In fact, \citet{gnedin04} found that the reheating from structure formation shocks significantly impacted their simulated 21cm absorption signal.  The sky-averaged 21cm brightness temperature, which LEDA and DARE aim to observe, is likely to be inversely proportional to the gas temperature at the time of emission, so that colder temperatures would result in a larger signal.  Here we show that structure formation shocks are not likely to suppress this absorption signal, in contrast to \citet{gnedin04}.

In addition, recently \citet{tseliakhovich10} argued that a previously unconsidered effect could impress large spatial fluctuations in the 21cm signal:  At the time of recombination, the cosmic gas was moving with respect to the dark matter at a root mean square velocity of $30~$km~s$^{-1}$ and in a coherent manner on $\lesssim10$~comoving~Mpc separations.  These initial velocity differences translate into the dark matter moving through the gas with a typical Mach number of $\calM \approx 2$ from $z = 150$ until the time the gas was reheated.  Semi-analytic models predict that astrophysical backgrounds reheated the Universe somewhere between $z=10$ and $20$ \citep{furlanetto06,pritchard10}.

The baryon-dark matter supersonic motion could suppress the formation of the first stars \citep{tseliakhovich10, maio11, stacy11, naoz11b}.  It could also introduce spatially variable kinetic temperatures by the entropy generated in shocks that these supersonic flows sourced.  Both of these effects would result in fluctuations in the 21cm signal.  In fact, the 21cm correlation function may have a significantly larger amplitude on the $10-100~$comoving~Mpc scales that observations are most sensitive to if this differential velocity impacts the 21cm background \citep{dalal10}.    Already, many 21cm models that do not account for the \citet{tseliakhovich10} effect find that the fluctuations in the 21cm background from $z=20$ are likely to be no more difficult to detect than those from $z=10$ \citep{pritchard08}.  Existing interferometric efforts to detect highly redshifted 21cm emission are primarily focused on $z= 6-10$.  If the signal from $z=20$ were further boosted by the dark matter--baryon velocity differential (a scenario the results of this paper support under specific circumstances), then more effort should be channeled towards detecting 21cm fluctuations from this earlier epoch.  Furthermore, the power spectrum of the differential velocity features significant acoustic features similar to those in the CMB.   These features could provide a more distinctive signature that would aid the separation of the 21cm signal from the foregrounds.  

The aim of this paper is to estimate the coupling strength of the 21cm signal to the dark matter--baryon velocity differential.  To do so, this study presents semi-analytic estimates for the size of this coupling, which are calibrated with numerical simulations of the Dark Ages.  Our simulations, described in detail in a companion paper \citep[Paper I]{paperI}, employ more physical cosmological initial conditions than the simulations used in prior studies.  Unlike previous studies, our simulations are initialized with a transfer function that consistently incorporates the dark matter--baryon differential velocity.  (On the scale of our simulations, $\leq 1~\cMpc/h$, this differential velocity is a uniform wind to an excellent approximation.)  In Paper I, we show that these improvements provide a much better match to the linear evolution of cosmological perturbations.  In addition, we run both the Enzo \citep{oshea04} and GADGET \citep{springel01} cosmological codes, with many different box sizes and particle numbers/grid sizes, to explore the robustness of our results. 


This paper is organized as follows:  Section \ref{sec:21cm} provides a short introduction into the 21cm signal, its detectability, and introduces a parameterization for how this signal could couple to the dark matter--baryon velocity differential.  We then provide estimates for how the 21cm signal is impacted by this velocity differential under various assumptions, using a suite of simulations to calibrate these assumptions (Section \ref{sec:estimates}). This study assumes a flat $\Lambda$CDM cosmological model with $\Omega_m=0.27$, $\Omega_\Lambda=0.73$, $h=0.71$, $\sigma_8= 0.8$, $n_s=0.96$, $Y_{\rm He} = 0.24$, and $\Omega_b = 0.046$, consistent with recent measurements \citep{larson11}.  We will subsequently abbreviate proper Mpc as pMpc and Mpc and kpc will be reserved for comoving lengths.  Some of our calculations use the Sheth-Tormen mass function, for which we adopt the parameters $p = 0.3$, $a=0.75$, $A=0.322$  \citep{sheth02}.\\

While this study was nearing completion, \citet{visbal12} published a related study using semi-numeric methods.  \citet{visbal12} investigated one model for how the baryon--dark matter velocity difference could impact the 21cm signal.  We provide here a more complete census of the different scenarios where the relative velocity could contribute to the 21cm signal.








\section{The Pre-reionization 21cm Signal}
\label{sec:21cm}

This section provides a brief introduction to the temporal evolution of the redshifted 21cm signal.  This introduction differs slightly from prior expositions (e.g., \citealt{furlanetto06, furlanettoohbriggs}) in that it characterizes the transition in terms of the star formation rate density, $\dot \rho_{\rm SFR}$, a quantity that is also constrained by optical/infrared observations of the $z\sim 10$ Universe.  One goal of \S \ref{ss:temporal} is to motivate why it is likely for the IGM to appear with $\Tk < \TCMB$ in 21cm absorption soon after the first stars formed (the situation this paper primarily investigates).  A second goal is to understand whether minihalos could source the pre-reionization radiation background, which is the scenario that leads to the differential velocity having its largest impact on the 21cm signal.  In addition, minihalo scenarios have largely (and we argue unjustly) been ignored in previous research on the extremely-redshifted 21cm background.   Next, \S \ref{ss:fluct} provides an introduction to the fluctuating 21cm signal and also motivates why the velocity difference between the baryons and dark matter could lead to larger fluctuations than in previous models that had neglected this effect.  Finally, \S \ref{sec:detectability} discusses the detectability of 21cm signals.

\subsection{the temporal evolution of this signal}
\label{ss:temporal}
The 21cm line of atomic hydrogen offers a brightness temperature contrast with respect to the CMB brightness temperature of
\begin{eqnarray}
T_b^{21\cm} &=& 41.5 \, x_{\rm H} \,(1+\delta_b) \left(1 - \frac{T_{\rm CMB}(z)}{\TS} \right)  Z_{20}^{1/2}{\rm ~ mK} ,\label{eqn:Tb1}\\ 
 &=& -216 \, x_{\rm H}  \,(1+\delta_b) \left( \frac{1 - \frac{T_{\rm CMB}}{\TS}}{1 -  \frac{T_{\rm CMB}}{T_{\rm K}^{\rm ad}}}\right) Z_{20}^{1/2} {\rm ~~ mK}, \label{eqn:Tb2}
\end{eqnarray}
where $T_{\rm CMB}(z)$ is the CMB temperature at redshift $z$, $Z_{20} \equiv (1+z)/21$, $\TS$ is the spin temperature of the 21cm line \citep{field58}, $x_{\rm H}$ is the neutral hydrogen fraction, and $\delta_b$ is the overdensity in baryons.  Note that for ensuing expressions we will sometimes drop the superscript ``$21\cm$'' in $T_b^{21\cm}$.
  Equation~(\ref{eqn:Tb2}) was evaluated for $\TS$ equal to the kinetic temperature of the gas prior to reheating and after thermal decoupling or $T_{\rm K}^{\rm ad} = 0.021 \,(1+z)^2~$K.  This equation demonstrates that the absolute brightness temperature of the 21cm signal is likely to peak at times when the gas was kinetically cold.

 
The spin temperature of the 21cm transition interpolates between the gas temperature, $\Tk$, and $\TCMB(z)$ as \citep{field58}
\begin{equation}
\TS^{-1} = \frac{\TCMB^{-1} + T_{\rm K}^{-1} (x_\alpha + x_c)}{1+ x_\alpha + x_c},
\label{eqn:Tspin}
\end{equation}
where $x_c$ describes how well particle collisions couple $\TS$ to $\Tk$.  This coefficient is effectively zero in the low-density IGM at $z<40$ (the redshifts we focus on), and 
\begin{equation}
x_\alpha \approx 1.8 \times 10^{11} \, (1+z)^{-1} J_\alpha
\label{eqn:xalpha}
\end{equation}
parametrizes the efficacy of Ly$\alpha$ scattering at pumping the 21cm transition.  This scattering couples $\TS$ to $\Tk$ via the Wouthuysen-Field mechanism  \citep{field58}.  Equation~(\ref{eqn:xalpha}) assumes c.g.s.\ units for $J_\alpha$, where $J_\alpha$ is the average photon specific intensity at the frequency of Ly$\alpha$ and is given by (e.g., \citealt{pritchard06})
\begin{eqnarray}
J_\alpha &=& \sum_{n=2}^{\infty} f_{\rm rec}(n) \int_z^{z_{\rm max}(n)} dz' \frac{(1+z)^2}{4\pi} \frac{c}{H(z)} \epsilon(\nu_n', z'),~~~~\\ \label{eqn:Jalpha}
           &\approx& \frac{5}{27}\frac{c  \,(1+z)^{3}}{4 \pi H(z)} \; \epsilon_*(z).
\label{eqn:approxJ}
\end{eqnarray}
Here, $H(z)$ is the Hubble expansion rate, $\epsilon$ is the spatially-averaged comoving specific emissivity, $f_{\rm rec}(n)$ is the probability that an absorption into the $n^{\rm th}$ Rydberg level of atomic hydrogen -- and the resulting radiative cascade -- produces a Ly$\alpha$ photon \citep{pritchard06}, $z_{\rm max}$ is the maximum redshift at which a photon could have been produced that was absorbed into the $n^{\rm th}$ level at $z$, and $\nu_n' = \nu_n (1+z')/(1+z)$.  The approximation that yields equation~(\ref{eqn:approxJ}) includes only the $n=2$ contribution to the summation:  Photons that redshift into the Ly$\beta$ resonance contribute negligibly to the pumping of $T_{\rm S}$ (these photons are destroyed after a few scatterings) and higher Lyman resonances than Ly$\alpha$ are pumped by photons in a more restricted spectral range.  Equation~(\ref{eqn:approxJ}) also makes the further simplification that the specific emissivity is frequency-independent between $1$ and $3~$Ry with value $\epsilon_*$.\\

The evolution of the 21cm brightness temperature is thought to be determined by the production of ultraviolet photons that pump the 21cm line (affecting $x_\alpha$), X-ray heating of the intergalactic gas ($T_{\rm K}$), and the reionization of hydrogen ($x_{\rm H}$).  In what follows, we discuss the order in which these different processes likely impacted the 21cm line:\\ \\
\noindent {\bf Ly$\alpha$ pumping:  }
Given equations (\ref{eqn:xalpha}), (\ref{eqn:Jalpha}), and the number of $0.75-1~$Ry photons produced per baryon incorporated in stars, $N_{\alpha}$, we can solve for the comoving star formation rate density required to satisfy $x_\alpha = 1$ and thereby couple $\TS$ to $\Tk$:
\begin{equation}
[\dot \rho_{\rm SFR}]_{\alpha} = 1.7 \times 10^{-3} \, Z_{20}^{-1/2} \left(\frac{N_{\alpha}}{10^4}\right)^{-1}~{\rm \Msun ~yr^{-1} ~\cMpc^{-3}}.\label{eqn:SFRalpha}
\end{equation}
Stellar population synthesis calculations find $N_{\alpha} \approx 10^4$ for Pop~II stars with a standard initial mass function (IMF; \citealt{leitherer99}) and $N_{\alpha}  \approx 5000$ for a top-heavy Pop~III stellar population \citep{bromm01}.   

Let us estimate the epoch at which $[\dot \rho_{\rm SFR}]_{\alpha}$ is surpassed, assuming that the fraction $f_\star$ of the baryons in halos above mass $m_h$ are incorporated into stars such that $\dot \rho_{\rm SFR} = f_\star \, \bar{\rho}_b \, df_{\rm coll}(m_h)/dt $.  Here, $f_{\rm coll}(m_h)$ is the fraction of matter that has collapsed in halos more massive than $m_h$, for which we use the Sheth-Tormen halo mass function.  For $f_\star = \{0.1\%, 1\%, 10\%\}$, the characteristic star formation rate, $[\dot \rho_{\rm SFR}]_{\alpha}$, would have been encountered at $\{6,~ 15,~ 19\}$  if star formation traced the mass in halos that could cool atomically.  These estimates assumed that such halos have virial temperatures of $T_{\rm vir} > 10^4~$K, yielding halo masses of $m_H > 3\times 10^7 \, Z_{20}^{-3/2}$.   

Most previous discussions of 21cm radiation have assumed that halos that cool atomically are the main driver of the Ly$\alpha$ pumping \citep{gnedin04, furlanetto06}.    However, stars in smaller ``minihalos'' (halos that can only cool by molecular hydrogen transitions) may contribute to the production of ultraviolet photons.  Section \ref{sec:estimates} will show that $\sim 10^6~\Msun$ halos \emph{must} generate much of the $0.75-1~$Ry photons for the differential dark matter--baryon differential velocity to modulate the local star formation rate in a detectable manner.  Minihalos with circular velocities of at least $3.7~$km~s$^{-1}$ are massive enough for the gas to be able to cool by molecular hydrogen \citep{tegmark97, abel02, machacek01}, and they correspond to halo masses of $> 3.5\times 10^5\, Z_{20}^{-3/2}~\Msun$.    Such halos would have achieved $x_\alpha = 1$ at $z = \{13,~ 22,~ 28\}$ for $f_\star = \{0.1\%, 1\%, 10\%\}$.  If instead the bulk of the star formation occurred in more robust $1.5\times 10^6\, Z_{20}^{-3/2}~\Msun$,  $6~$km~s$^{-1}$ halos, $x_\alpha = 1$ would have occurred at somewhat lower redshifts of $z = \{11, 20, 25\}$.

The Lyman-Werner band background (e.g., the spectral band $11.2-13.6$~eV) acts to destroy the molecular hydrogen and sterilize star formation, especially in the lowest mass minihalos \citep{haiman97, haiman00}.  The numerical calculations of \citet{machacek01}, which were confirmed in \citet{wise07}, found that $4\%$ of the gas is able to cool (and collapse to much higher densities) in halos with mass
\begin{equation}
m_{\rm crit} = 2.5 \times 10^{5} + 8.7 \times 10^{5} \, F_{LW, 21}^{0.47}~~~\Msun,
\label{eqn:mcLW}
\end{equation}
where $F_{LW, 21}$ is the Lyman-Werner intensity integrated over solid angle in units of $10^{-21} ~{\rm erg~s^{-1}~Hz^{-1}~sr^{-1}}$.\footnote{\citet{oshea08} found a similar relation, but with a $\sim 3$ times higher normalization.}
The critical star formation rate to raise $m_{\rm crit}$ turns out to be much less than $[\dot \rho_{\rm SFR}]_{\alpha} $:  $x_\alpha = 1$ requires a radiation background in this band with intensity of $F_{\alpha, 21} = 4 \pi h \nu_{\rm {\rm Ly} \alpha}\,(1+z)/ (1.8 \times 10^{11})\times10^{21}$ in the same units as $F_{LW, 21}$  (e.g., eqn. \ref{eqn:xalpha}).  Plugging in the numbers, $x_\alpha = 1$ yields a Lyman-Werner intensity of
\begin{equation}
F_{LW, 21} \approx 20 \,Z_{20} \, \exp[-\tau_{LW}],
\label{eqn:FLW}
\end{equation}
noting that $F_{LW, 21} \sim F_{\alpha, 21}$ and $\tau_{LW}$ is the typical intergalactic opacity for Lyman-Werner photons which can can be $1-2$ in the absence of dissociations \citep{ricotti01} and can be larger once the first HII regions have formed \citep{johnson07}.  Once $x_\alpha = 1$ and assuming equation (\ref{eqn:mcLW}), the Lyman-Werner background is sufficient to suppress cooling in halos with masses $<4 \times 10^6 \exp[-0.47 \tau_{LW}]~\Msun$.\footnote{In addition, if an appreciable number of solar mass stars form, these produce infrared radiation that dissociates H$^-$, preventing the formation of molecular hydrogen. \citet{wolcott12} found that such an infrared background becomes more effective at preventing star formation than the Lyman-Werner background if $\sim$solar mass stars comprise $>90\%$ of the stellar mass.  This would require the high-redshift IMF to be more bottom heavy than the $z=0$ IMF.}

\citet{wise08} studied the star formation rate density in cosmological simulations that followed Pop~III star formation in minihalos, including self-consistently Lyman-Werner radiation backgrounds.  They found that star formation rate densities sufficient for $x_\alpha =1$ occurred by $z= 22-25$ in their simulations. \\

\noindent {\bf X-ray heating:  }
Another critical juncture in the evolution of the 21cm signal occurred when $\rhoSFR$ was sufficient for X-rays to have heated the gas above the CMB temperature.  Penetrating X-rays are likely the most efficient mechanism for reheating the IGM \citep{chen04}.  However, relating X-ray production to the star formation rate (SFR) is more uncertain than relating ultraviolet emission to SFR since X-ray production depends on the abundances of X-ray binaries and supernovae.  To do so, we follow the methodology taken in \citet{furlanetto06}, using relations calibrated on low-redshift galaxies between X-ray luminosity and the SFR.  In particular, the critical SFR density to heat the IGM with X-rays by an amount of $T_{\rm CMB}(z)$ is
\begin{eqnarray}
[\dot \rho_{\rm SFR}]_{\rm X} &=& 4.0\times 10^{-2}  \, Z_{20}^{5/2}\,\left( \frac{t_{\rm SFR}}{0.1 \,t_H} \right)^{-1} \, \left(\frac{f_X}{0.2} \,  \right)^{-1} \label{SFRX}  \\
& \times & \left(\frac{L_X/{\rm SFR}}{10^{40} {\rm ~erg~ s^{-1}~ \Msun^{-1} ~yr}} \right)^{-1} ~{\rm \Msun ~yr^{-1} ~Mpc^{-3}}, \nonumber 
\end{eqnarray}
where $f_X$ is the fraction of energy that heats the IGM \citep{shull85}, $L_X/{\rm SFR}$ is the $\sim0.1-2~$keV luminosity per unit SFR, $t_{\rm SFR}$ is the timescale over which the emitting population had been active, and $t_H = H(z)^{-1}$.  Equation~(\ref{SFRX}) evaluated $L_X/{\rm SFR}$ at $10^{40}  {\rm ~erg~ s^{-1}~ \Msun^{-1} ~yr}$, which is a factor of $\sim5$ higher than low redshift measurements for the same relationship except between $2-10~$keV \citep{grimm03, mineo11}.\footnote{The spectral index of the X-ray emission is uncertain, but empirical determinations at low-redshifts are consistent with having equal energy per log in frequency \citep{rephaeli95, swartz04}.  Low-redshift X-ray emission that traces star formation is dominated by high-mass X-ray binaries.}  Interestingly, there is no evidence for evolution in the $L_X/{\rm SFR}$, even to $z\approx 6$ \citep{cowie11}.  Using the Sheth-Tormen mass function, the $e$-folding time, $[d \log f_{\rm coll}(m_h)/dt]^{-1}$, is $0.11$ and $0.06$ of a Hubble time at $z=20$ for $m_h =10^6~\Msun$ and $m_h =10^8~\Msun$, respectively.

Is it possible for minihalos to have also dominated the X-ray reheating of the Universe?
For our fiducial parameters, $F_{LW, 21}$ would have been even larger by a factor of $20$ when $[\dot \rho_{\rm SFR}]_{\rm X}$ was satisfied than when $[\dot \rho_{\rm SFR}]_{\alpha}$ was.  Larger $f_X$ will reduce the resulting $F_{LW, 21}$, making it more difficult for the Lyman-Werner background to sterilize minihalos.  However, X-rays can also catalyze the formation of $H_2$, combatting its destruction via the Lyman-Werner background \citep{haiman97}.  We find that at the epoch when $[\dot \rho_{\rm SFR}]_{\rm X}$ is satisfied the electron fraction is increased by a factor of $10$ over the relic electron fraction at the cosmic mean density.  A $10$ times higher electron fraction means that $10$ times more molecular hydrogen will form, such that a $10$ times larger Lyman-Werner background is required to yield the same $H_2$ fraction.  However, this estimate  represents an upper bound on how much the critical $F_{LW, 21}$ of a halo would be increased by $X$-rays, as the recombination time is $\sim 1$ Hubble time for $\delta_b = 200$ gas with the relic electron fraction at $z=20$.  Thus, it is difficult to significantly boost the electron fraction in dense, star forming regions with X-rays.  In fact, \citet{machacek03} found in simulations that an X-ray background only mildly promotes the formation of molecular hydrogen in halos.
\\


\noindent {\bf Ionizations:  }
The final effect that stars have on the $21$cm signal is via their ionizations.  If a stellar population produces $N_{\rm ion}$ hydrogen ionizing photons per stellar baryon, the critical SFR density required to reionize the Universe to an ionized fraction of $x_i$ is
\begin{eqnarray}
[\dot \rho_{\rm SFR}]_{\rm ion} &=  & 4.4 \times 10^{-1} \, \bar x_i \, Z_{20}^{3/2} \,\left( \frac{t_{\rm SFR}}{0.1 \,t_H} \right)^{-1}  \nonumber \\
&  & \times \left(\frac{f_{\rm esc}}{0.1}  \frac{N_{\rm ion}}{4000}\right)^{-1} ~{\rm \Msun ~yr^{-1} ~Mpc^{-3}},\label{eqn:RI}
\end{eqnarray} 
where $f_{\rm esc}$ is the fraction of ionizing photons that escape from their sites of production into the IGM.  The factor $f_{\rm esc}$ is highly uncertain (e.g., \citealt{kuhlen12}) and likely to be $\ll 1$.  For Pop~II stars with a Scalo initial mass function (IMF), $N_{\rm ion} \approx 4000$ for $Z = 0.05 \, Z_\sun$  \citep{barkana01}.  This number varies at the factor of $2$-level when changing assumptions regarding the metallicity and the IMF, at least for empirically-determined IMFs.  However, Pop~III stars with a top-heavy IMF are much more efficient producers of ionizing photons, with $N_{\rm ion} \approx 40, 000$  \citep{bromm01}.\\  

\begin{figure}
{\epsfig{file=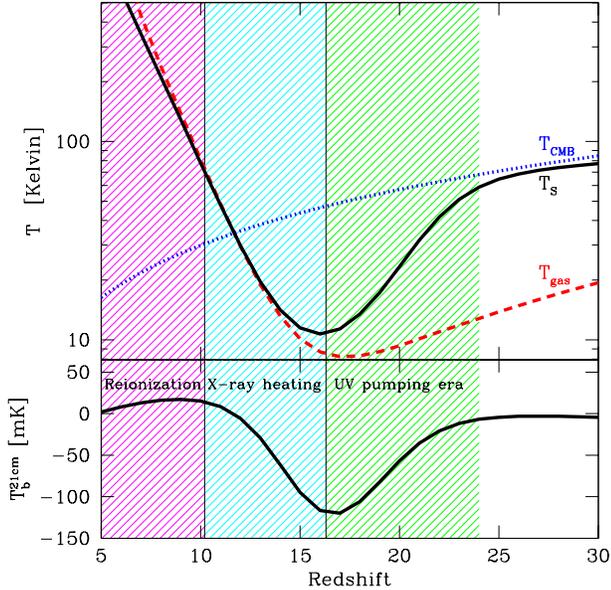, width=8.5cm}}
\caption{Model history of the gas temperature and spin temperature (top panel) and of the mean 21cm brightness temperature (bottom panel) for the parameters $N_\alpha = 10^4$, $f_X = 1$, $f_* = 0.02$, $N_{\rm ion} = 4000$, and $f_{\rm esc} = 0.1$, assuming that star formation traces the mass in atomic cooling halos.  The shaded regions qualitatively delineate the phases where different radiation backgrounds drive the signal:  first radiative pumping by far ultraviolet photons, then heating by penetrating X-ray ionizations, and lastly reionization by extreme ultraviolet photons.  The LEDA and DARE instruments aim to constrain this signal between $10 \lesssim z \lesssim 30$.   \label{fig:global}}
\end{figure}

Thus, we find $[\dot \rho_{\rm SFR}]_{\alpha} \ll [\dot \rho_{\rm SFR}]_{\rm X} \ll [\dot \rho_{\rm SFR}]_{\rm ion}$ in agreement with the progression found in \citet{furlanetto06}.  With this ordering, radiation from star formation first coupled the spin temperature to the gas temperature such that the 21cm signal would appear in absorption.  Next, radiation associated with star formation reheated the Universe, and, lastly, ultraviolet radiation from stars reionized the cosmic gas.  
 In fact, \citet{pritchard10} demonstrated that one could change both $N_\alpha$ and $f_X$ by orders of magnitude and still have a phase in which the 21cm signal was in absorption owing to Ly$\alpha$ coupling $\TS$ to $\Tk$ at times prior to X-ray reheating above $\TCMB$.  The amplitude of the 21cm signal is larger during this phase compared to its amplitude at other times (eqn.~\ref{eqn:Tb2}). The discussion in this paper focuses on this phase.

 Figure \ref{fig:global} shows the history evolution of $\Tk$, $\TS$ and the 21cm brightness temperature for the parameters $f_X = 1$, $f_* = 0.02$, and $f_{\rm esc} = 0.1$ (defined in eqn.s \ref{eqn:SFRalpha}, \ref{SFRX}, and \ref{eqn:RI}), and assuming that star formation traces atomic cooling halos.   The shaded regions qualitatively delineate the phases where different processes drive the signal.  In this model, the 21cm signal appears in absorption until $z\approx 11.5$ (bottom panel, Fig. \ref{fig:global}).  The temperature history for a model in which minihalos dominate the backgrounds can be similar, but the minihalo case is more likely to be important as redshift increases.
 
Our discussion has ignored shock heating as a contribution to the gas temperature.  The Universe could heat owing to structure formation shocks.  Such heating would also suppress the absorption signal that occurs at $z \approx 12-22$ in Figure \ref{fig:global}.  Shock heating would be unfortunate as this dip (the sharpest and strongest feature that is expected in the sky-averaged 21cm signal) is being targeted by the 21cm instruments LEDA and DARE.  Its presence is also important for the effects studied in this paper.  In the only previous numerical study of this dip's existence, \citet{gnedin04} found that it was significantly impacted by shock heating.  Fortunately, our simulations (which are better suited for this calculation than those in \citealt{gnedin04}) show that shock heating does not significantly impact the temperature of the intergalactic gas at these redshifts (see Appendix \ref{ap:shocks}).

Relating the critical $\rhoSFR$ derived in this section to the $\rhoSFR$ measured in rest-frame ultraviolet observations is also helpful for gauging when these different $\rhoSFR$ thresholds are satisfied.  In particular, \citet{bouwens10} found $\rhoSFR \approx 10^{-2}~\Msun~\cMpc^{-3}~{\rm yr}^{-1}$ at $z=8$ as well as that  $\rhoSFR$ was decreasing quickly with increasing redshift.  A star formation rate density of $10^{-2}~\Msun~\cMpc^{-3}~{\rm yr}^{-1}$ is only just sufficient to reionize the Universe in one Hubble time for $f_{\rm esc} =0.5$ -- a value that is higher than anticipated --, but it is more than sufficient to satisfy $[\dot \rho_{\rm SFR}]_{\alpha}$.  At face value, the rest-frame ultraviolet determinations of $\rhoSFR$ also suggest that the $\rhoSFR$ thresholds outlined in this section are satisfied at relatively low redshifts.  However, it is likely that faint, dwarf galaxies that are missed by the \citet{bouwens10} observations contribute significantly to the true value of $\rhoSFR$ at $z\gtrsim8$ \citep{bouwens11, kuhlen12}.



\subsection{Fluctuations}
\label{ss:fluct}

Interferometric efforts targeting redshifted 21cm radiation are not directly sensitive to the mean $21$cm signal, but instead to spatial fluctuations in the $21$cm brightness temperature.  
In the limit of small fluctuations (appropriate for the models we consider), the observed brightness temperature contrast with respect to $\TCMB$ can be approximated as (e.g., \citealt{furlanettoohbriggs})
\begin{eqnarray}
\delta T_b^{21\cm} \approx ~~\bar{T}_b^{21\cm} \, &\bigg(& 1+\delta_b + \frac{1}{1+\bar{x}_\alpha}\delta_\alpha \nonumber \\
& +&  \frac{T_{\rm CMB}}{\bar{T}_{\rm K} - T_{\rm CMB}}\delta_T - \delta_{\grad v} \bigg),
\label{eqn:deltaT}
\end{eqnarray}
where $\bar{T}_b^{21\cm}$ is the spatial average of equation (\ref{eqn:Tb1}) and bars over other quantities also denote a spatial average.  Equation (\ref{eqn:deltaT}) is valid at $z \lesssim 40$, once collisions can no longer pump the $21$cm line, and prior to reionization.   Respectively, $\delta_b$, $\delta_\alpha$,  $\delta_T$, and $\delta_{\grad v}$ are the overdensities in baryons, Ly$\alpha$ radiation, temperature, and the line-of-sight derivative of the line-of-sight proper velocity (arising from redshift-space distortions).  


At lowest order, the overdensities in the temperature and in the intensity at the Ly$\alpha$ resonance can be related to the overdensity in baryons and the square of the baryon--dark matter velocity difference, $v_{\rm bc}^2$, as
\begin{eqnarray}
\delta_\alpha &=& \left(b_{\delta, \alpha}\, \delta_b + b_{v^2, \alpha} \, \delta_{v^2}  \right) \star W_\alpha, \label{delta_alpha}\\
\delta_T &=& \left(b_{\delta, T}\, \delta_b + b_{v^2, T} \, \delta_{v^2}  \right),
\label{delta_T}
\end{eqnarray}
where 
\begin{equation}
\delta_{v^2} \equiv \left[v_{\rm bc}^2(\bfx)/\sigma_{\rm bc}^2 - 1\right].
\end{equation}
\citet{dalal10} demonstrated that the expansion to first order in $\delta_{v^2}$ is a good approximation if $\delta_\alpha$ and $\delta_T$ are local functions of $|\bfv_{\rm bc}|$, at least on scales where fluctuations in $v_\bc^2$ are much less than unity.
In addition, $b_{\delta, \alpha}$ is the bias of star forming regions, $b_{\delta, T}$ is the bias of density-tracing temperature fluctuations (equal to $2/3$ for adiabatic evolution)\footnote{Both temperature ``biases'' formally enter in convolution at times when X-ray heating was important, reflecting the propagation of X-ray photons.} and $b_{v^2, \alpha}$ [$b_{v^2, \alpha}$] is the bias with which star formation [heating] traces $\delta_{v^2}$.  Lastly, $\sigma_\bc^2 \equiv \langle v_{\rm bc}^2 \rangle_v$, where 
$\langle ... \rangle_v$ denotes an ensemble average over the Maxwell-Boltzmann probability distribution of $v_{\rm bc}$.
   In addition, $W_\alpha(r)$ describes how Ly$\alpha$ photons travel from a point source to a distance $r$ away and is normalized to have unit norm, entering equation~(\ref{delta_alpha}) in convolution.  It is given by a similar expression to equation~(\ref{eqn:approxJ}) \citep[see][]{barkana05}.

Interferometric 21cm fluctuation efforts aim to measure the power spectrum of the 21cm brightness temperature fluctuations, $P_{21}(\bfk)  = \langle |\delta \tilde T_b(\bfk)|^2 \rangle$, where tildes signify the Fourier dual.\footnote{Any significant contribution to the power spectrum from $v_{\rm bc}$ also means the signal is highly non-Gaussian such that higher-order statistics such as the trispectrum will be easier to detect than previously anticipated \citep{yoo11}.  However, we concentrate on the power spectrum here.}  Combining equations~(\ref{eqn:deltaT}), (\ref{delta_alpha}), and (\ref{delta_T}) yields
\begin{eqnarray}
& &\bar{T}_b^{-2} P_{21}(\bfk) \approx  \nonumber \\
 && ~~~\left(1+  \, \frac{b_{\delta, \alpha}\, \tilde{W}_\alpha(k)}{1+\bar{x}_\alpha} +   \frac{T_{\rm CMB} \, b_{\delta, T}}{\bar{T}_{\rm K} - T_{\rm CMB}} + \mu^2 \right)^2 P_\delta(k) \nonumber \\
& &~~~ +  \left(\frac{b_{v^2, \alpha}\, \tilde{W}_\alpha(k)}{1+\bar{x}_\alpha}  + \frac{T_{\rm CMB} \,b_{v^2, T}}{\bar{T}_{\rm K}  - T_{\rm CMB}} \right)^2 P_{v^2}(k), \label{eqn:P21} 
\end{eqnarray}
where $P_\delta \equiv \langle |\tilde \delta(k)|^2 \rangle$ and $P_{v^2}\equiv \langle |\tilde \delta_{v^2}(k)|^2 \rangle$.  We are working in the limit in which the baryons trace the matter such that $\delta_b = \delta$, and we have dropped the much smaller terms that are $\propto \langle \tilde \delta(k)^2  \tilde \delta_{v^2}(k) \rangle$, even though they formally are the same order as $P_{v^2}$.  The terms with $\mu^2$ owe to redshift-space distortions \citep{kaiser87}, where $\mu = \nhat \cdot \bfk/k$ and $\nhat$ is a unit vector that points along the line-of-sight.   

Previous analyses aside from \citet{dalal10} had not included the $P_{v^2}$ contribution to $P_{21}$.  This new term's amplitude peaks at smaller $k$ than $P_\delta$.  In fact, even if the coefficient that couples it to $P_{21}$ is $10^4$ times smaller than the analogous coefficient for $P_\delta$, it still would contribute comparable power to this other term at $k = 0.1~$Mpc$^{-1}$ -- roughly the scale at which 21cm experiments are most sensitive \citep{mcquinn06, parsons11}.  The Fourier transform of $P_{v^2}$ is equal to
\begin{equation}
\left \langle \delta_{v^2}(\bfx) \, \delta_{v^2} (\bfx +\bfr) \right \rangle  = \frac{4}{9} \, \psi_1(r)^2  +  \frac{2}{9}\, \left[\psi_1(r) + \psi_2(r) \right]^2 ,
\label{eqn:xivsq}
\end{equation}
where
\begin{eqnarray}
\psi_1(r) &\equiv & \frac{3}{\sigma_\bc^2}\int_0^\infty \frac{k^2 dk}{2 \pi^2} P_{v}(k)  \frac{j_1(k r)}{k r},\\
\psi_2(r) &\equiv & \frac{3}{\sigma_\bc^2}\int_0^\infty \frac{k^2 dk}{2 \pi^2} P_{v}(k) \ j_2(k r),\\
\sigma_\bc^2 &=& \int_0^\infty \frac{k^2 dk}{2 \pi^2} P_{v}(k) ,
\label{eqn:sigmabc}
\end{eqnarray}
$P_{v} \equiv \langle |\tilde{v}_{\rm bc}(\bfk)|^2\rangle$, and
\begin{equation}
\tilde{v}_{\rm bc}(\bfk) = - i \frac{a \, \bfk}{k^2} \left[\dot{T_c}(k, a) - \dot{T_b}(k, a)\right] \tilde \delta_{\rm pri}.
\label{ref:vbck}
\end{equation}
A dot over a variable signifies differentiation with respect to time.  $T_c$ and $T_b$ are the baryonic and dark matter transfer functions that map from the primordial overdensity, $\delta_{\rm pri }$, to the overdensity in these components.  Equation~(\ref{eqn:xivsq}) can be derived from equation~(\ref{ref:vbck}) using Wick's theorem and the correlation function between the cartesian components of the velocity field:
\begin{equation}
\left \langle v_i(\bfx)  v_j(\bfx+\bfr) \right \rangle_v = \frac{\sigma_\bc^2}{3} \, \left( \psi_1(r) \, \delta_{ij}^{\rm K} + \psi_2(r) \, \frac{r_i r_j}{r^2} \right),
\end{equation}
where $\delta_{ij}^{\rm K}$ is the Kronecker delta.

One of the primary objectives of this paper is to calculate the contribution of the term proportional to $P_{v^2}$ to $P_{21}$, namely to calculate $b_{v^2,\alpha}$ and $b_{v^2, T}$ in equation (\ref{eqn:P21}).\\


\noindent {\bf An estimate for $b_{v^2,\alpha}$:  }
If star formation traces the amount of matter that collapses into halos above some mass cutoff, $f_{\rm coll}$, \citet{dalal10} found that
\begin{equation}
b_{v^2, \alpha}  \approx  \frac{3}{2} \left(\frac{\langle v_{\rm bc}^2 f_{\rm coll} \rangle_v}{ \sigma_\bc^2 \langle  f_{\rm coll} \rangle_v} -1 \right),
\label{eqn:bvalpha1}
\end{equation}
provided a good fit to the large-scale bias found in a full calculation for the source clustering in the model they considered.  
An alternative bias coefficient comes from equating the variance in $f_{\rm coll}$ as a function of $v_\bc$ to $\langle \delta_{v^2}^2 \rangle_v$ (which equals $2/3$), and is given by
\begin{equation}
b_{v^2, \alpha} \approx \sqrt{\frac{3}{2}}\, \left( \frac{\langle f_{\rm coll}^2 \rangle_v}{\langle f_{\rm coll} \rangle_v^2} -1 \right)^{1/2}.
\label{eqn:bvalpha2}
\end{equation} 
This alternative bias should hold to the extent that the $v_{\rm bc}$--driven fluctuations are linear on \emph{all} scales.  Such a condition does not hold for the density field at $z\lesssim 20$, but is more valid for $\delta_{v^2}$.

The exercise of plugging $f_{\rm coll}(v_\bc) \rightarrow \langle f_{\rm coll} \rangle_v \, (1+ b_{v^2, \alpha} \, \delta_{v^2})$ into the right-hand-side of equations~(\ref{eqn:bvalpha1}) and (\ref{eqn:bvalpha2}) reveals that both expressions for the bias are exact in the limit that higher than linear order terms are subdominant.  We find that either choice of bias agrees within $\sim 10\%$ among the models we consider.  However, because the calculations in \citet{dalal10} find that the bias given by equation~(\ref{eqn:bvalpha1}) fares excellently even in the somewhat nonlinear cases they considered, our subsequent calculations will use equation~(\ref{eqn:bvalpha1}) for $b_{v^2, \alpha}$ and for $b_{v^2, T}$ when radiation backgrounds source the fluctuations.\\

\noindent {\bf An estimate for $b_{v^2,T}$:  }
We derive $b_{v^2, T}$ with a similar logic to how equation~(\ref{eqn:bvalpha2}) was derived:  If temperature-overdensity fluctuations were coupled to $v_{\rm bc}^2$ with root mean square (RMS) fluctuation amplitude of $\sigma_T$ when averaging over regions, then
\begin{equation}
b_{v^2, T} = \sqrt{\frac{3}{2}} \, \sigma_T,
\label{eqn:bsigT}
\end{equation}
in order for the variance of  the field $b_{v^2, T} \, \delta_{v^2}$ to equal $\sigma_T$.
As an example of heating, if the fraction $\beta$ of $v_{\rm bc}^2$ were locally thermalized in shocks (such that the region was heated by the factor $1+ 5\beta \calM^2/9$), 
\begin{equation}
b_{v^2, T} =  \frac{5 \beta}{9} \, \left \langle \calM_{\rm bc}^2 \right \rangle_v / \left(1+\frac{5\beta}{9} \left \langle \calM_{\rm bc}^2 \right \rangle_v \right).
\label{eqn:bs}
\end{equation}


\subsection{detectability}
\label{sec:detectability}

It is not necessarily more difficult to detect the fluctuating 21cm signal from the era of focus here, $z\sim 20$, than from $z \approx 8$ -- the epoch most efforts are targeting:  The absolute brightness temperature of the signal is likely to be much larger at $z\sim 20$ than at $z\sim 8$, which compensates for the increased brightness temperature of the foregrounds \citep{pritchard08}.  

For subsequent estimates of the 21cm fluctuating signal, we quantify its potential to be detected by considering an interferometric configuration similar to the Murchison Widefield array (MWA) but that is optimized to the higher redshifts in question (by dilating the sizes of the dipoles and baselines).  We also consider an interferometer with $10$ times the collecting area that is a scaled up replica of the MWA.  See Appendix \ref{ap:sensitivity} for additional details regarding these hypothetical instruments.
The statistical sensitivities of these instruments (excluding sample variance) to $P_{21}$ measured in spherical shells of width $\Delta k = 0.2 \,k$ are shown in the ensuing 21cm power spectrum figures.  These calculations assume an observing time of $1000~$hr and a bandwidth of $5~$MHz.  They also assume that the foregrounds can be removed from all wavevectors that do not utilize the zero mode along the line-of-sight such that the sensitivity at these wavevectors is limited by thermal noise \citep{mcquinn06}.  For $B = 5~$MHz, an instrument can at best observe $k>0.05~$Mpc$^{-1}$, corresponding to the fundamental line-of-sight wavenumber for the chosen bandwidth.


Larger bandwidths (which allow a measurement at smaller $k$) are desirable to probe the $k$ where $P_{v^2}$ has its largest impact on the $P_{21}(k)$.
However, the bandwidth (redshift) interval must also be chosen so that there is not significant temporal evolution in the signal across the band.   The amount of collapsed mass evolves on a timescale of $\Delta z/(1+z) \sim 0.1$ (Section \ref{ss:temporal}), which is comparable at $z=20$ to the $\Delta z$ covered by a bandwidth of $5~$MHz.  Thus, we expect the amount of star formation evolves substantially across the redshifts covered in this instrumental band.  This evolution would result in spurious large-scale power at the $[T_b^{\rm 21cm}]^2$--level, which dwarfs the signal we are discussing.  Thus, extracting the $v_{\rm bc}$--induced signal from the lowest $k$ to which a survey is sensitive will be challenging (which happen to be where the effects of $v_{\rm bc}$ are most prominent).



\section{The impact of $v_{\rm bc}$ on the 21cm signal}
\label{sec:estimates}

\begin{figure}
\rotatebox{-90}{
\epsfig{file=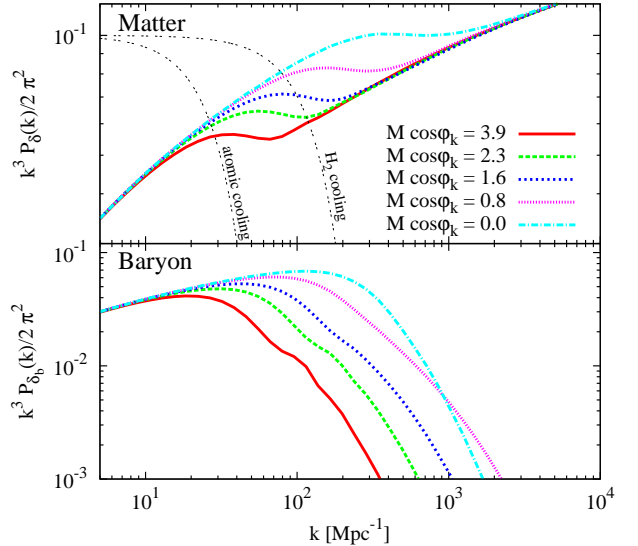, width=7.7cm}}
\caption{Impact of $v_{\rm bc}$ on the total matter and baryonic power spectrum at $z=20$.  Shown are modes in which ${\calbfM}_\bc \cdot \hat \bfk$ equals the quoted number.  The top panel shows the total matter power spectrum, and the bottom panel shows the power spectrum of just the baryons.  For reference, 95\% of space has $\calM_\bc > 0.8$, 70\% has $\calM_\bc > 1.6$, and $5\%$ has $\calM_\bc > 3.9$.  Also shown in the top panel is the squared Fourier transform of the top-hat in real-space window function (normalized to $0.1$ at $k=0$) that in the spherical collapse model would collapse into halos with circular velocities of $10~$km~s$^{-1}$ and $3.7~$km~s$^{-1}$ -- the thresholds at which atomic and molecular hydrogen cooling become important, respectively.  \label{fig:velEffect}}
\end{figure}

This section provides analytic estimates for the imprint of $v_{\rm bc}$ on the 21cm signal.  These estimates are calibrated with the cosmological simulations discussed in Paper I.  
They also require a calculation for how $v_{\rm bc}$ alters the linear growth of density fluctuations.  To do so, we solve the linear system of equations given in Paper I, and originally in \citet{tseliakhovich10}, for the growth of modes in the presence of a nonzero $v_{\rm bc}$.  This calculation is initialized with $z=1000$ transfer functions from the CAMB code.\footnote{\url{http://camb.info/}}  It assumes an instantaneous kinetic decoupling of the baryons from the CMB at this redshift (as motivated in \S~5.3.1 in \citealt{hu95} and \citealt{eisenstein98}) to solve for the growth of structure in two gravitationally coupled fluids, the dark matter and the baryons.  This approximation for the growth of modes avoids the full Boltzmann code calculation.  We find that solutions with this approximation excellently reproduce the growth of modes in CAMB for the case CAMB solves, $v_{\rm bc} = 0$. 

Figure \ref{fig:velEffect} features the results of this calculation for $z=20$, showing the small-scale power spectrum of the total matter (top panel) and the baryons (bottom panel).  A nonzero $v_{\rm bc}$ acts as an effective pressure in the baryons, decreasing the $k$ above which pressure smooths fluctuations and in an anisotropic fashion.  The linear variance in the baryonic density contrast is altered at the $\sim 20 \%$ level at $z=20$ by $v_{\rm bc}$.  Nonlinear evolution further amplifies these differences.

\subsection{temperature fluctuations from structure formation and shocking}
\label{ss:sf}

\begin{figure}
\rotatebox{-90}{
\epsfig{file=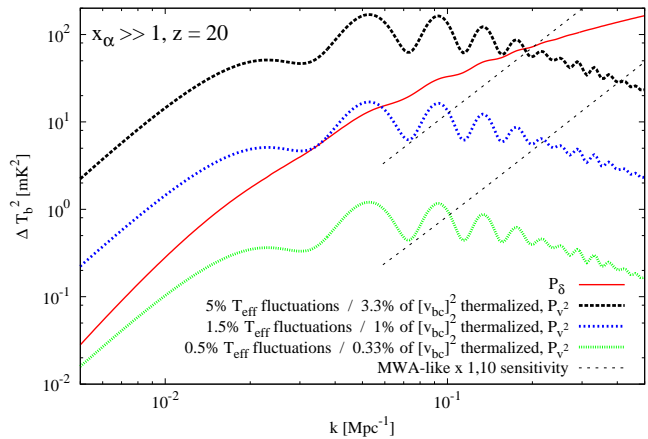, width=6.2cm}}
\caption{$P_{v^2}$ and $P_{\delta}$ contributions to the azimuthally-averaged 21cm power spectrum, where $\Delta T_b(k)^2 \equiv (2 \pi^2)^{-1} k^3 P_{21}$, in cases where the effective gas temperature, $T_{\rm eff} \equiv \langle 1/\Tk \rangle_M^{-1}$, is modulated by the local value of $v_{\rm bc}$.  The different curves represent the case where the standard deviation in temperature, $\sigma_T$, is the specified value or where the specified fraction of $v_{\rm bc}^2$ is thermalized.   While the curves are for $z=20$ and $x_\alpha \gg 1$, the relative amplitudes of the $P_{v^2}$ curves are redshift independent, and their overall amplitude scales as $\bar{T}_b^{2}/(\bar{T}_{\rm K} - \TCMB)^2$.  The amplitude of the $P_{\delta}$ curve scales with the additional factor of $(1+z)^{-2}$.  The diagonal lines are the statistical sensitivity to the power spectrum in bins of $\Delta k = 0.2 \,k$ for an array that has $1$ and $10$ times the number of antenna elements as MWA, assuming an observing time of $1000~$hr and $B = 5~$MHz.  \label{fig:Pkheating}}
\end{figure}

It is possible that shocking or structure formation changed the temperature of the Universe between regions with different $v_\bc$ sufficiently to result in an enhancement of the 21cm signal.  The period when such heating is most likely to source a significant component in the 21cm signal is after Ly$\alpha$ pumping started to couple the spin temperature to the temperature of the gas but before the gas was reheated above the temperature of the CMB by X-rays.  Figure \ref{fig:Pkheating} illustrates how the 21cm power spectrum, $\Delta T_b(k)^2  \equiv k^3 P_{21}/ (2 \pi^2)$, would be affected if $0.33$, $1$, and $3.3\%$ of the energy in $v_{\rm bc}$ were thermalized using equation~(\ref{eqn:bs}) and assuming $x_\alpha \gg 1$.  (The relative amplitude of the curves do not depend on $x_\alpha$.)  If just a small fraction of the kinetic energy in $v_\bc$ were thermalized, this heating would lead to large temperature fluctuations and a larger 21cm signal than has been anticipated in models that ignore $v_\bc$.  We showed in Paper I that the gaseous dynamical friction timescale is shorter than the Hubble time for dark matter overdensities with masses greater than $10^4~\Msun$ -- roughly halos massive enough to significantly perturb the gas density.  The short dynamical friction time results in the streaming gas decelerating into the dark matter potential wells and in visually apparent shocks throughout the simulation volume (see Fig.~4, in Paper I).  In fact, by $z=20$ in the simulation, all gas with $\delta_b > 2$ has lost its relative velocity owing to dynamical friction.  The amount of energy dissipated in this deceleration is enough to significantly heat the cosmic gas (by a factor of $3$, on average) if it were all thermalized.  However, it is difficult to estimate from first principles the amount of energy that this process would actually thermalize:  This heating may primarily impact overdense gas, which occupies a small fraction of the cosmic volume, and the low Mach number shocks sourced by $v_\bc$ are inefficient entropy generators.\footnote{
\noindent  For low Mach number shocks, the energy lost in deceleration may not necessarily go into heating the gas.  The temperature increase across an adiabatic shock is 
\begin{equation}
\frac{T}{T_0} = 16^{-1} \, (5 \calM^2 - 1)(3 \calM^{-2} +1) ~~\overrightarrow{_{~as~\calM \rightarrow 1^+~}}~~ \calM.
\end{equation}
More fundamental is the entropy jump, where we define entropy here as $S \equiv p/\rho^{5/3}$.  In this case,
\begin{equation}
\frac{S}{S_0} = 4^{-8/3} \, (5 \calM^2 - 1)(3 \calM^{-2} +1)^{5/3} ~~\overrightarrow{_{~as~\calM \rightarrow 1^+~}} ~~\frac{5 (\calM-1)^3}{6},
\end{equation}
which is zero at first and second order in $\calM - 1$, where $\calM$ is the shock Mach number.  
For an $\calM = 2$ shock, $\{T/T_0, S/S_0\}$  equals $\{2.1, 1.2\}$ and, for an $\calM = 4$, it equals $\{5.9, 2.6\}$.  Thus, the regions where $\calM_\bc$ is largest may lead to the most heating (even though dynamical friction is less effective there).
In addition, if the simulations fail to capture $\calM-1 = 0.1$ shocks, they are missing only $\approx 10^{-3}$ jumps in entropy.
}  


Finite $v_{\rm bc}$ suppresses the collapse of gas onto halos, which also impacts the gas temperature.  If the suppression of structure leads to temperature fluctuations between regions with different $v_{\rm bc}$ that have standard deviations of $\sigma_T = 0.5\%$, $1.5\%$, or $5\%$, each would lead to a contribution to $P_{21}$ given by one of the $P_{v^2}$ curves in Figure \ref{fig:Pkheating}.  We can estimate the level of temperature fluctuations that result from the impact of $v_\bc$ on structure formation.  In the limit that $T_b^{21\cm} \propto \rho/\Tk$ (that applies when $x_\alpha \gg 1$) and pure adiabatic evolution such that $\Tk \propto (1+\delta_b)^{2/3}$, the average 21cm brightness temperature depends on $v_{\rm bc}$ as
\begin{eqnarray}
 \bar{T}_b^{21}  (v_{\rm bc}) &=& \bar T_{b}^{21} \,\left \langle \left(1+ \delta_b \right)^{1/3} \right \rangle_{{\rm fixed}~v_\bc}, \nonumber \\ 
 &\approx& \bar T_{b}^{21}   \left( 1 - \frac{1}{9} \, \sigma_{\delta_b}(v_{\rm bc})^2 + ... \right),
 \label{eqn:Tstruct}
\end{eqnarray}
where $\sigma_\delta(v_{\rm bc})$ is the standard deviation in the gas density in a region with differential velocity $v_{\rm bc}$.  Thus, the average temperature is slightly lower in regions with larger variance, at least when $\sigma_{\delta_b} \ll 1$ so that higher order moments are not important. 
Computing $\sigma_T = \langle \bar{T}_b^{21}  (v_{\rm bc})/ \langle T_{b}^{21} \rangle_v -1 \rangle_v^{1/2}$ using linear theory yields $0.002$ at $z=20$ (a time when the linear theory standard deviation in $\delta_b$ is $\approx 0.5$).  At $z=20$ and for $x_\alpha \gg 1$, the imprint of $\sigma_T = 0.002$ on the 21cm power spectrum would lie a factor of $6$ below the $0.5\%$ curve in Figure \ref{fig:Pkheating}, and, thus, be quite small. 

Equation (\ref{eqn:Tstruct}) neglects the impact of Compton heating, which results in deviations from pure adiabatic evolution.  Compton heating changes the temperature of a region by $0.6~K~$per Hubble time at $z\approx 20$ for the cosmic free electron fraction of $2\times10^{-3}$, although the heating rate scales as $(1+z)^4$.  Because the Compton heating rate is density independent (noting that the electron fraction is largely uniform) whereas the adiabatic cooling rate scales as $\delta^{2/3}$, the fractional temperature in voids is more impacted by Compton heating than in filaments.  This results in a $v_\bc$--dependence to the heating as larger $v_\bc$ results in shallower void depths. 



\begin{figure}
\rotatebox{-90}{\epsfig{file=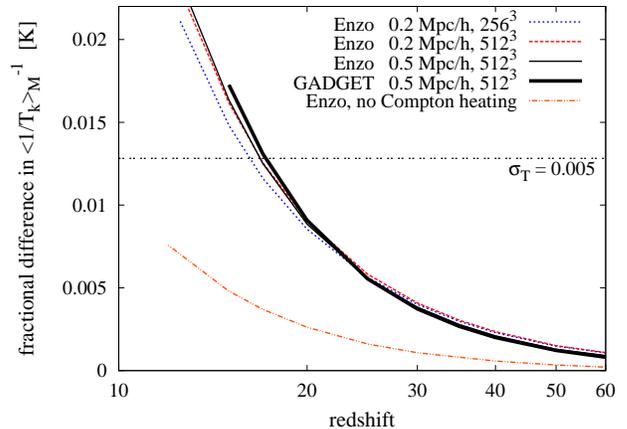, width=6cm}}
\caption{Fractional difference between the 21cm intensity--weighted temperature in simulations with $\calM_{\bc} = 0$ and in those with $\calM_{\bc} = 1.9$ (corresponding to $v_\bc = 3$~km~s$^{-1}$ at $z=100$).  Note that the temperature in the simulations with $\calM_{\bc} = 0$ is higher, and the fractional temperature difference is similar in Enzo to this difference in GADGET.  The horizontal line shows the difference that leads to the signal given by the lowermost $P_{v^2}$ curve in Figure~\ref{fig:Pkheating}.   The Enzo simulations here are uni-grid, but we find negligible differences when compared with the Enzo AMR calculations. \label{fig:Tdiff}}
\end{figure}

Thus, shocks, structure formation and Compton heating result in $v_\bc$--dependent temperatures.  Cosmological simulations are our best recourse for a more accurate estimate. We use the cosmological codes GADGET3
\citep{springel01} and Enzo v2.1.1 \citep{oshea04} with initial conditions that include $v_\bc$ self-consistently (in addition to other improvements; Paper I).  GADGET solves the equations of fluid dynamics with the smooth particle hydrodynamics method, whereas Enzo is a grid code with adaptive mesh refinement (AMR).    Both the GADGET and Enzo codes have been shown to
conserve entropy at the part in $1000$--level for the test case of the expansion of a homogeneous Universe \citep{oshea05}.  This level of entropy conservation is unusual for hydrodynamics codes, and it owes to the
entropy-conserving formalism of GADGET and the $3^{\rm rd}$--order
accurate in space, $2^{\rm nd}$--order in time Riemann solver employed
by Enzo.  Thus, both codes are well-motivated choices for tracking thermal effects in the early Universe, and their much different hydrodynamics solvers tests the robustness of the results.\footnote{In addition, we have put the cosmological codes GADGET and Enzo
 through a battery of tests in order to confirm that our results
 regarding the temperature evolution of the Universe are robust.  We find that the relative difference in the 21cm intensity--weighted
 temperature in the simulations is robust to: i) the frame of reference for
 the relative velocity of the baryons and dark matter on the grid at least for Galilean transforms with boost velocity comparable to $M_\bc$ ii) the
 maximum time-step size, iii) the grid size, iv) the box size, v) the number of
 particles, vi) the number of AMR levels in Enzo, and vii) the chosen hydrodynamics solver in Enzo.}  Each curve in Figure  \ref{fig:Tdiff} plots the difference in $\langle 1/\Tk \rangle_M^{-1}$ between both GADGET and Enzo simulations between a simulation with $\calM_{\bc} = 0$ and one with $\calM_{\bc} = 1.9$, where $\langle ... \rangle_M$ signifies a mass average.  We refer to this difference as $\delta \langle 1/\Tk \rangle_M^{-1}$.    Note that $\bar T^{21}_{b} \propto [1+ \TCMB \, \langle \Tk^{-1} \rangle_M]$ when $x_\alpha \, \TCMB/\Tk \gg 1$ such that $\langle \Tk^{-1} \rangle_M^{-1}$ is the gas temperature--weighting relevant to 21cm observations during the period of interest (when this signal appears in absorption).

All of our simulations find that the average temperature is larger in the simulations with $\calM_{\bc} = 0$ than in those with larger $\calM_{\bc}$.  (Note that on the scale of our simulations, $\leq 1~\cMpc/h$, this differential velocity is a uniform wind with a single Mach number, allowing us to refer to simulations by their value of $\calM_{\bc}$.)  The direction of the temperature difference indicates that the
suppression of structure formation and the impact of Compton heat-
ing are the dominant processes that shape the temperature rather than shock heating.  The dot-dashed curve shows $\delta \langle 1/\Tk \rangle_M^{-1}$ in Enzo between a simulation with and one without Compton heating. Compton heating contributes more than half of this temperature difference.   


With both Enzo and GADGET, the temperatures are reasonably converged in resolution and in box size between the simulations shown in Figure \ref{fig:Tdiff}.\footnote{This is not the case if we run GADGET in the standard setting rather than with the gravitational softening equal to the gas softening as done here, where we find the temperature differences are twice as large owing to artificial particle coupling.}  We also find that Enzo is converged in $\delta \langle 1/\Tk \rangle_M^{-1}$ between both the adaptive-mesh refinement (AMR) and uni-grid simulations.  We find that the temperature differences are roughly linear in $v_{\bc}$ from comparing with also the $\calM_{\bc} = 3.8$ simulations such that we find $\sigma_T \approx 0.4 \, \delta \langle 1/\Tk \rangle_M^{-1}$.  This allows us to draw a horizontal line in Figure \ref{fig:Pkheating} that corresponds to the lowest amplitude curve for the $P_{v^2}$ contribution to $P_{21}$ in Figure \ref{fig:Tdiff}.  For $v_{\bc}$--sourced temperature fluctuations to contribute a comparable fraction of the 21cm power as the density sourced intensity fluctuations, $\sigma_T \gtrsim 0.015$ must be satisfied (Fig. \ref{fig:Pkheating}).   Thus, the simulations yield temperature differences that are a factor of a few too small for the $P_{v^2}$ contribution to be an ${\cal O} (1)$ contribution to $P_{21}$.

\subsection{star formation--sourced fluctuations}
\label{ss:SFR}


While we found that the coupling of non-astrophysical processes to $v_\bc$ is insufficient to significantly enhance the fluctuations in the 21cm background, the local amount of star formation can also couple to $v_\bc$ and modulate associated radiation backgrounds, which in turn affect $P_{21}$.  We attempt to model this coupling here.  We consider two disparate models for which halos at $z\sim 10-20$ dominated the SFR:
\begin{description}
\item[Molecular Cooling Halos:]  In this case, stars are primarily formed in minihalos with $10^{2.5} \lesssim T_{\rm vir} \lesssim 10^4~K$.  These halos cool by exciting molecular hydrogen transitions.  We use a simplistic parametrization to quantify how $v_\bc$ could impact these halos.  
In particular, in addition to including the impact of $v_\bc$ on the halo mass function, we assume that all halos with higher circular velocities than $V_{\rm cool} (z)$ are able to retain their gas and form stars proportional to their mass.  As in \citet{fialkov11}, we use 
\begin{equation}
 V_{\rm cool} (z) = \{V_{\rm cool, 0}^2 + \left [\alpha \, v_{\rm bc}(z) \right]^2\}^{1/2}.\label{eqn:alpha}
\end{equation}
Previous numerical simulations find $V_{\rm cool,0} = 3.7 ~{\rm km ~s}^{-1}$ -- the approximate threshold to cool via $H_2$ when $F_{\rm LW, 21} \ll 1$ and in the absence of baryonic streaming -- and $\alpha = 4.0$ (e.g., \citealt{fialkov11}, who used fits to the numerical results of \citealt{stacy11} and \citealt{greif11}).  
We leave $\alpha$ as a free parameter that is calibrated with our simulations, and we use $V_{\rm cool,0} = 3.7 ~{\rm km ~s}^{-1}$ as well as $V_{\rm cool,0} = 7.4~{\rm km ~s}^{-1}$.  The latter choice roughly corresponds to the minimum circular velocity that can cool for a Lyman-Werner background with $x_\alpha = 1$ (eqn. \ref{eqn:mcLW} evaluated with the intensity from eqn. \ref{eqn:FLW}).  For reference, $3.7 ~{\rm km ~s}^{-1}$ and $7.4~{\rm km ~s}^{-1}$ correspond to $4\times10^5~\Msun$ and $3\times 10^6~\Msun$ halos at $z=20$.  
\item[Atomic Cooling Halos:]  In this case, halos with $T_{\rm vir} > 10^4~$K dominate the total star formation rate (e.g., \citealt{gnedin04, furlanetto06}).  It is unlikely that $v_\bc$ significantly impacts the mass threshold that cools and forms stars in this case because (1) atomic transitions are a robust coolant that depends more weakly on the density of the gas and (2) the circular velocity of these halos is much larger than $v_\bc$.   The impact of $v_\bc$ on star formation in these halos should derive primarily from its effect on these halos' mass function.  However, Figure \ref{fig:velEffect} illustrates that $v_\bc$ has a small impact on the matter power spectrum at scales relevant to atomic cooling halos, so a large $v_\bc$-sourced contribution to $P_{21}$ is not anticipated.
\end{description}
For both the molecular and atomic cooling cases, our calculations assume that star formation is proportional to collapsed mass above the minimum mass threshold.


\subsection{numerical estimates for the star formation suppression}

\begin{figure}
\rotatebox{-90}{
\epsfig{file=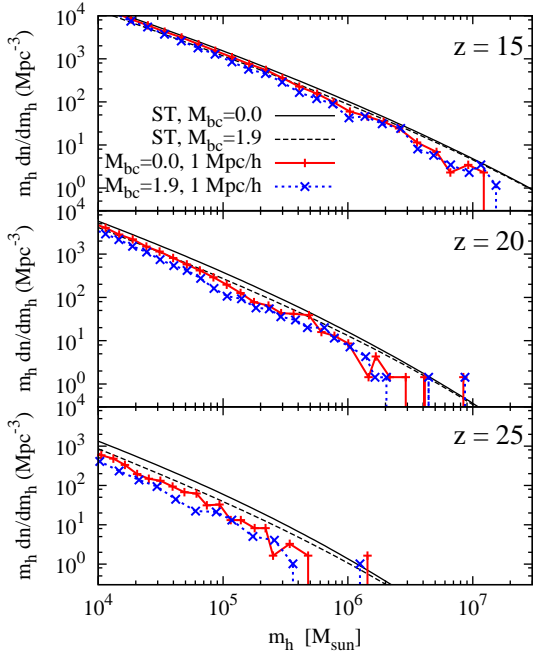, width=9cm}}
\caption{Impact of $v_\bc$ on halo dark matter mass function for the $1~\cMpc/h$, $2\times768^3$ particle GADGET simulations (the largest simulations we ran).  Also shown is the Sheth-Tormen mass function for the same $v_\bc$ calculated for a region that has zero overdensity on the box scale, as described in the text, to compare on an equal footing. \label{fig:mf}}
\end{figure}

Ultimately, our aim in this section is to estimate the contribution to the 21cm power spectrum that traces $P_{v^2}$ from star formation--sourced radiation backgrounds.  We use our cosmological simulations to motivate how the star formation rate is suppressed.  
We first investigate $v_\bc$'s impact on the halo mass function in numerical simulations relative to what one would predict for its impact on the Sheth-Tormen halo mass function (which can be calculated directly from the linear theory matter power spectrum).  Each panel in Figure \ref{fig:mf} shows halo mass functions calculated from our two largest GADGET simulations ($1~\cMpc/h$, $2\times768^3$ particle, one with $\calM_\bc =0$ and the other with $\calM_\bc=1.9$).\footnote{In Paper I we showed that particle coupling can impact the GADGET simulations when $\calM_\bc =0$ and this particle coupling is alleviated in the simulations with $\calM_\bc >0$.}  These mass functions were calculated using the friends-of-friends algorithm with a linking length of $0.2$.  For comparison, we also show the semi-analytic mass function using the Sheth-Tormen theory.  We include finite box size effects in the Sheth-Tormen mass function as described in the ensuing footnote in order to compare on an equal footing with the simulation mass functions.\footnote{To account for finite box size effects in Figures \ref{fig:mf} and \ref{fig:SFRhist}, we multiply the Sheth-Tormen mass function by $n_{\rm PS}(m_h | \sqrt{\sigma_{m_h}^2 - \sigma_{l_{\rm box}/2}^2})/ n_{\rm PS}(m_h | \sigma_m)$, where $n_{\rm PS} (m_h, \sigma_X)$ is the Press-Schechter mass function at mass $m_h$ given $\sigma_X$, the RMS density contrast in a sphere of radius X.  This prescription was motivated in \citet{barkana04}.}  Both sets of curves are calculated with $\calM_\bc$ equal to $0$ and $1.9$, and both simulations use the same random numbers to generate the density and velocity fields.  While the simulated mass function is slightly below the Sheth-Tormen mass function, particularly at $z=25$, the suppression of the amplitude of the mass function is comparable to that predicted by the semi-analytic mass function calculation.  We will use this semi-analytic mass function model in subsequent calculations.

\begin{figure}
\begin{center}
\epsfig{file=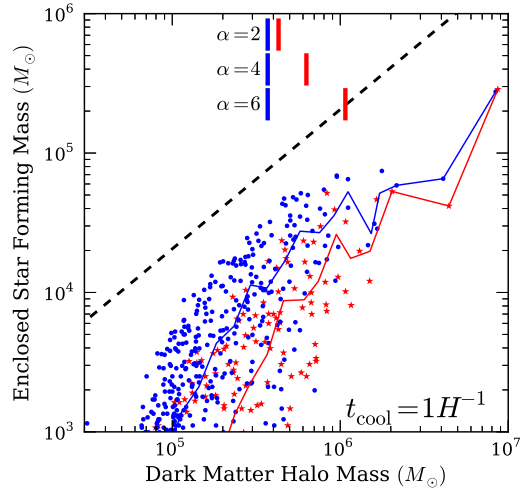, width=7cm}
\end{center}
\caption{Impact of $v_\bc$ on mass that can cool within $1$ Hubble time as a function of halo mass, where each marker shows the gas mass that meets our criterion within a virial radius from a halo centroid.  The blue circles are the $\{1~\cMpc/h,  2\times768^3~{\rm particle}\}$ simulation with $\calM_\bc = 0$, and the red stars are the same but for $\calM_\bc = 1.9$.  The thin curves with the corresponding color show the mean in both simulations.   The dashed curve is the cosmic baryon fraction times the halo mass, and the short vertical lines illustrate the cutoff mass in our simple model for $\calM_\bc = 0$ and $\calM_\bc = 1.9$ and the specified $\alpha$.  Crudely, the impact of $\calM_\bc$ is to shift the minimum halo mass with $\alpha \approx 4$ in equation~(\ref{eqn:alpha}), but $\calM_\bc$ also suppresses star formation in more massive systems and in a stochastic manner.  \label{fig:SFRsup}}
\end{figure}

\begin{figure}
\begin{center}
\epsfig{file=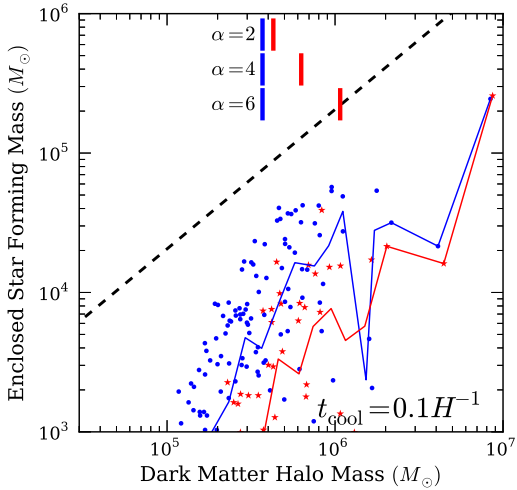, width=7cm}
\end{center}
\caption{Same as Figure \ref{fig:SFRsup} but for the amount of gas that can cool in $0.1$ Hubble times.
\label{fig:SFRsup2}}
\end{figure}

The other parameter we aim to calibrate from the simulations is $\alpha$.  This parameter regulates the minimum mass of a halo that is able to cool by molecular transitions.  Figures~\ref{fig:SFRsup} and \ref{fig:SFRsup2} show the mass in gas within a virial radius of a halo of the specified dark matter mass that can cool by molecular hydrogen transitions in $1.0$ and $0.1$ Hubble times, respectively, at $z=20$.  The points show the individual halos in the $1~\cMpc/h$, $2\times768^3~$particle simulations, and the curves are the average amount of mass in gas that can cool in these simulations.  To calculate the cooling time, we use the formula in \citet{tegmark97} under the crude approximation that $\log(1+ N_{\rm rec}) =1$ to calculate the amount of molecular hydrogen, where $N_{\rm rec}$ is the number of recombinations.  This equality approximately holds for gas at the virial density of halos at $z\sim20$, and it allows us to calculate the cooling rate from a single simulation snapshot rather than following the density evolution of a fluid element.\footnote{Most other studies of the impact of $v_\bc$ have followed the collapse of gas parcels to much higher densities as a proxy for star formation \citep{maio11, stacy11, greif11}.  
 In our opinion, it is not necessarily a disadvantage to use our cooling criteria on as a proxy for star formation rather than following the cooling and condensing gas to much higher densities:  Firstly, simulations that follow gas to much higher densities do not all agree on the character of star formation in the first halos (e.g., \citealt{greif11b}).  In addition, feedback processes either from stellar HII regions within the halo \citep{alvarez06, yoshida07} or the cosmological Lyman-Werner background \citep{haiman00, machacek01} drastically increase the complexity of modeling star formation at the Cosmic Dawn.  However, we have run the same simulations in Enzo with molecular hydrogen cooling and AMR, and found that our simple estimates did roughly reproduce the amount of gas that cooled to much higher densities.}

The amount of mass that can cool trends to zero at halo masses less than  $\approx 1-4\times10^5~\Msun$ (Fig.s~\ref{fig:SFRsup} and \ref{fig:SFRsup2}).  Such a threshold was anticipated from more detailed first star calculations (e.g., \citealt{machacek01}), and larger $\calM_\bc$ shift this turn-over mass to higher values.  The vertical lines show how much this mass would shift under our parametrization for the impact on molecular hydrogen cooling halos for $V_{\rm cool, 0} = 3.7~$km~s$^{-1}$ and the specified $\alpha$.  The shift in the critical mass that can form stars in the simulations most closely approximates the shift seen in the $\alpha =4$ case.

However, the $V_{\rm cool, 0}-\alpha$ parametrization does not explain the halo-to-halo stochasticity in the simulations.
There are far more halos in the $\calM_\bc =1.9$ simulation that have zero gas under our cooling criteria than in the $\calM_\bc =0$ simulation.  
A small component of this suppression results from the halos being slightly less massive in the $\calM_\bc =1.9$ case (see Fig. \ref{fig:mf}), but most originates from $v_\bc$'s suppression of gas accretion onto these halos.   In Paper I, we showed that one source of this variation from halo to halo results from the orientation of filamentary flows onto a halo. 
\begin{figure}
\begin{center}
\rotatebox{-90}{\epsfig{file=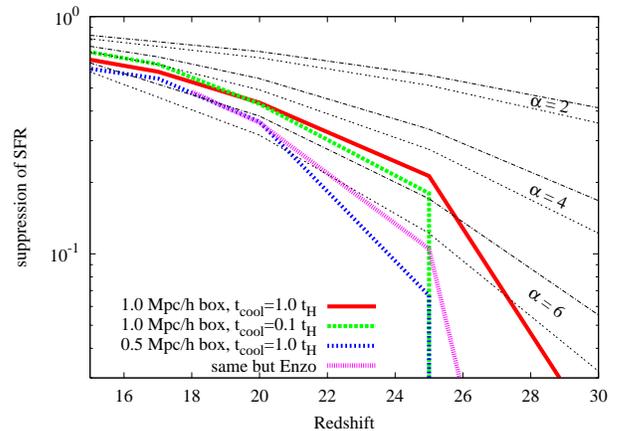, width=6cm}}
\end{center}
\caption{Ratio of the amount of gas that can cool within the specified period and form stars in simulations with $\calM_\bc = 0$ and $\calM_\bc = 1.9$.  Other than the one Enzo curve, all others are calculated from simulations run with GADGET.  Also shown for comparison are semi-analytic calculations for this ratio using the molecular hydrogen cooling model for the specified $\alpha$ for a box size of $1~\cMpc/h$ (thin dot-dashed curve) and $0.5~\cMpc/h$ (thin double dotted curve).  The simulations are closest to the model for $\alpha \approx 4-6$.\label{fig:SFRhist}}
\end{figure}

Figure \ref{fig:SFRhist} shows the ratio of the total gas that can cool in the simulation boxes with $\calM_\bc = 1.9$ to their counterpart with $\calM_\bc = 0$, under the same cooling criteria and using three sets of the GADGET simulations and one set of Enzo simulations.  Also shown for comparison are semi-analytic calculations for this ratio in the cases with $V_{\rm cool,0} =3.7~$km~s$^{-1}$ and with $\alpha =2$, $\alpha=4$ or $\alpha = 6$.  The simulation curves are most consistent with the $\alpha\approx 4-6$ curves, but note that the simulations become less reliable at the highest $z$ owing to the rareness of these halos.  We have also looked at the suppression of $\calM_\bc = 3.8$ and reached a similar constraint on $\alpha$. 

In conclusion, the factor by which star formation is suppressed is on average similar for both the $1~H^{-1}$ and $0.1~H^{-1}$ cases in Figures~\ref{fig:SFRsup}, \ref{fig:SFRsup2}, and $\ref{fig:SFRhist}$.  We note that the $0.1$ Hubble times criterion is likely most appropriate given the exponential growth with time in the number of collapsing halos.  Also, we find a similar suppression in our simulations with Enzo as with GADGET.   We conclude that $\alpha = 4-6$ is the best match to the suppression in the SFR under this simple parametrization.  While much of our discussion focussed on the $V_{\rm cool,0} =3.7~$km~s$^{-1}$, it seems natural to assume $\alpha$ is fixed as different physics changes $V_{\rm cool,0}$, such as the presence of a Lyman-Werner background.  We showed that this was the case if we changed the cooling time threshold (compare Fig.s~\ref{fig:SFRsup} and \ref{fig:SFRsup2}).  The ensuing discussion will further motivate this assumption.



\subsubsection{explanation for SFR suppression}
\label{sec:suppression}

We have found that the simulations prefer $\alpha \approx 4-6$ in equation (\ref{eqn:alpha}), which parameterizes the minimum star forming halo.  Let us attempt to understand this result with two toy models.
 An estimate for the overdensity of gas in a virialized minihalo when $v_\bc = 0$ and ignoring cooling is 
 \begin{equation}
 \delta_b(x) \approx  \left ( \frac{6 \,T_{\rm vir}}{5 \, T_{\rm K}^{\rm ad}} \right )^{3/2},
 \label{eqn:denshalo}
 \end{equation}
 where $ T_{\rm K}^{\rm ad}$ is the temperature of the IGM at virialization under the idealization of a purely adiabatic collapse and hydrostatic equilibrium \citep{tegmark97}.\footnote{Equation (\ref{eqn:denshalo}) is starting to break down at interesting $T_{\rm vir}$:  When $T_{\rm vir} >260$, equation (\ref{eqn:denshalo}) yields  $\delta_b > 200$ at $z=20$ (i.e., the densities it returns are higher than the virial density).}   In the case of finite $v_\bc$ and where its energy thermalizes in shocks during the collapse, this expression is altered such that $T_{\rm IGM, ad} \rightarrow  T_{\rm IGM, ad} (1 +5 \calM_\bc^2/9)$, resulting in  $\delta_b \propto (1 + 5\calM_\bc^2/9)^{-3/2}$.  There is some threshold virial temperature, $T_{\rm vir, *}$, that obtains high enough $\delta_b$ and $T_{\rm vir}$ to cool within a Hubble time.  The molecular-hydrogen cooling time is proportional to $\exp[512 {\rm ~K}/T]/n_{H_2}$.   At a crude level, we can ignore the factor $\exp[512 {\rm ~K}/T]$ because $V_{\rm cool, 0} = 3.7~{\rm km~s}^{-1}$ corresponds to an $800~$K halo at $z=20$ so that this factor is changing by less than the factor of $2$ above this velocity scale.  With this approximation, $T_{\rm vir, *}$ (and also $V_{\rm cool}^2$) scales as $\approx (1 + 5\calM_\bc^2/9)$ in this model, since $n_{H_2} \sim \delta_b$.  This scaling for $V_{\rm cool}$ results in a modulation comparable to $\alpha=6$ at $z=20$ in the parameterization in equation~(\ref{eqn:alpha}) and with $\alpha$ decreasing with increasing redshift.  (Including the factor $\exp[512 {\rm ~K}/T]$ suppresses the effective $\alpha$ somewhat.)

 An even simpler (but related) model uses that velocities scale as $(1+\delta)^{1/3}$ in an adiabatically collapsing region, at least in the absence of dissipative processes.  Thus, at halo densities $\delta \sim 200$, the streaming velocity accelerates and becomes $\approx 6 \, v_\bc$, where $v_\bc$ is the velocity at the cosmic mean density.  The total effective pressure at a halo virial radius scales as $\sim \{V_{\rm cool, 0}^2 + \left [6 \, v_{\rm bc}(z) \right]^2/3\}$, which has a similar form to the parameterization for $V_{\rm cool}$ given in equation (\ref{eqn:alpha}) and suggests $\alpha \approx 4$.  If the second term is larger than the first, the ram pressure is larger than the thermal pressure (potential depth).  One can also understand this scaling as the requirement that the circular velocity of the halo must be larger than the local streaming velocity  ($\sim 6 \, v_{\rm bc}$) for the gas to be focussed into the potential well, where it can shock and cool.

\subsubsection{Ultraviolet pumping}
\label{ss:Lyalpha}

\begin{figure}
\rotatebox{-90}{ \epsfig{file=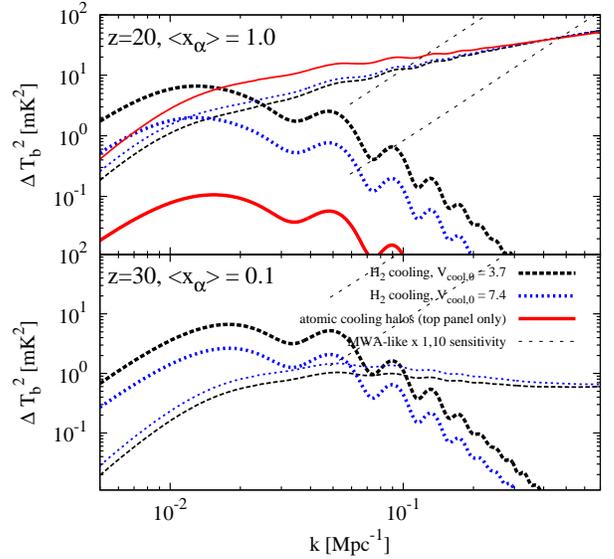, width=7.7cm}}
\caption{$P_{v^2}$ (thick curves) and $P_{\delta}$ (thin curves) contributions to the azimuthally-averaged $21$cm power spectrum for $\alpha = 4$, where $\Delta T_b(k)^2 =k^3 P_{21}/ (2 \pi^2)$, for different models of how $v_{\rm bc}$ modulates the ultraviolet pumping.  The top panel examines the case \{$z=20$, $\bar x_\alpha = 1$\} and the bottom \{$z=30$, $\bar x_\alpha = 0.1$\}.  See the text for details regarding the plotted models.  
  The sensitivity curves are for an MWA-like instrument but optimized for $z= 20$ and for a scaled-up MWA with ten times the collecting area.  These curves assume the same observing parameters as described in the caption of Figure \ref{fig:Pkheating}.  For $\alpha =6$, the minihalo curves that are sourced by $P_{v^2}$ shift upward by a factor of $\approx2$. \label{fig:Pk21}}
\end{figure}

Now that we have calibrated our model, finding $\alpha \approx 4-6$, we can use it to make predictions for the 21cm signal.  We first focus on 21cm inhomogeneities resulting form ultraviolet radiation from the first stars as this is likely to occur earlier than other star formation--sourced inhomogeneities (as argued in Section \ref{ss:temporal}).

Figure \ref{fig:Pk21} shows the $P_{v^2}$ and $P_{\delta}$ contributions to the $21$cm power spectrum, where $\Delta T_b(k)^2 \equiv k^3 P_{21}/ (2 \pi^2)$, for different models of how $v_{\rm bc}$ impacts star formation assuming \{$z=20$, $\bar x_\alpha = 1$\} in the top panel and \{$z=30$, $\bar x_\alpha = 0.1$\} in the bottom panel.  These curves assume that the X-ray heating and ionizations from these stars has yet to impart significant fluctuations in the 21cm background -- that the only fluctuation sources are density and $x_\alpha$.   Each curve can be rescaled to a different $\bar x_\alpha$ by noting that this parameter only changes the amplitude of $P_{21}$ via the factor $\bar{T}_b^2 (1+ \bar{x}_\alpha)^{-2}$. (This is not the case at $k \gtrsim 0.1~\cMpc^{-1}$ for the $P_\delta$ contribution, where the fluctuations begin to be sourced directly by $\delta_b$ rather than $J_\alpha$ as at smaller wavenumbers.) Our calculations also assume that the sources' specific emissivity scales as $\epsilon \propto \nu^\alpha$ with $\alpha = 0$, $\epsilon(z) = \exp[- (\tau_{\rm fcoll} H)^{-1} \, \Delta z/(1+z)]$, noting that $\tau_{\rm fcoll}^{-1} = d \log f_{\rm coll}(m_h)/dt$, $dz/(1+z) = H\,dt$, and $n_{\rm max} = 20$ (see \S \ref{ss:temporal}).\footnote{
We have ignored that that Lyman-Werner background can suppress star formation in regions that produce that are producing Ly$\alpha$.  \citet{holzbauer12} found that such fluctuations are unlikely to have a large impact on $P_{21}$.  It would decrease the amplitude of both the $P_{v^2}$ and $P_\delta$--coupled 21cm fluctuations arising from star formation.\\
\indent ~~~~~~In addition, we have not included the effects of lensing and peculiar velocities on $\delta_\alpha$ and $\delta_T$.  These effects enter at $1/b$ \citep{barkana05} and alter the contribution proportional to $\delta_b$ at the $\sim10\%$ level (but do not quantitatively effect our results).  Both effects do not alter the component that traces $P_{v^2}$.
}  We assume the same parametrization for  $\epsilon(z)$ when considering X-ray heating in \S \ref{sec:Xray}.

  The thick curves in Figure \ref{fig:Pk21} are the $P_{v^2}$ contribution to $P_{21}$ and the thin are this but for $P_\delta$.  The molecular cooling halo cases with $\alpha=4$ are given by the dashed curves ($V_{\rm cool, 0} > 3.7~{\rm km~s}^{-1}$) and dotted curves ($V_{\rm cool, 0} > 7.4~{\rm km~s}^{-1}$).  The same calculations but with $\alpha=6$ shift the $P_{v^2}$ curves in the minihalo--sourced models upward by a factor of $\approx 2$.  The atomic cooling halos case is given by the solid curves, and it only appears in the top panel since atomic cooling halos would not exist in sufficient abundance to pump the 21cm line at $z=30$.  In all cases, the relative impact of $v_\bc$ becomes smaller with time (decreasing redshift) because (1) $P_\delta$ is growing as $(1+z)^{-2}$, (2) halos at all mass scales are becoming less rare such that the impact on the growth of modes is less accentuated, and (3) $v_\bc$ and, hence, its role as an effective pressure for a halo with potential depth $\sim V_{\rm cool, 0}^2$ is becoming smaller with time.

Thus, we find in both considered minihalo cases that a significant component of the power is sourced by $P_{v^2}$ at $k\sim 10^{-2}~\cMpc^{-1}$ (Fig. \ref{fig:Pk21}).  This is consistent with the findings of \citet{dalal10}, although our model for the impact of $v_\bc$ on star formation is more conservative.  Unfortunately, the power is suppressed at larger wavenumbers by the free streaming of ultraviolet photons, and 21cm instruments are most able to observe $k \gtrsim 0.1~\cMpc^{-1}$ for reasons detailed in \S \ref{sec:detectability}.   Thus, we conclude that $P_{v^2}$ does not contribute significantly to $P_{21}$ on \emph{observable} scales at $z=20$.  For the $z=30$ case in Figure \ref{fig:Pk21}, the $P_{v^2}$ contribution is beginning to become comparable to the contribution from $P_\delta$ at $k\approx 0.1~\cMpc^{-1}$, roughly the largest observable scale.  However, observing the signal at $z=30$ is a more challenging venture.

We can also quantify how much both (1) $v_\bc$'s effect on the suppression of the growth of dark matter structure and (2) its impact on disrupting gas accretion onto minihalos (and modulating the minimum mass that can form stars) contribute to the inhomogeneities $v_\bc$ sources in the 21cm power spectrum.  We find that effect (2) tends to dominate in our models, sourcing $\approx 80\%$ of the $P_{v^2}$--coupled signal at $z=20$ in both minihalo models.  This percentile also holds for our ensuing estimates in \S \ref{sec:Xray}.


\subsubsection{X-ray heating}
\label{sec:Xray}

\begin{figure}
\rotatebox{-90}{
\epsfig{file=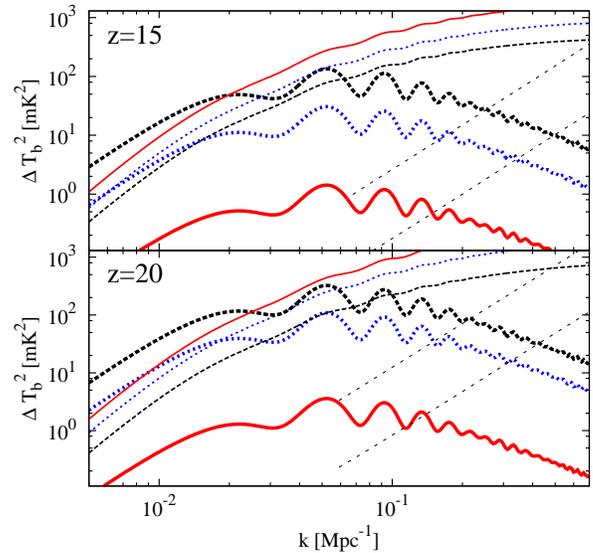, width=7.7cm}}
\caption{$P_{v^2}$ (thick curves) $P_\delta$ (thin curves) contributions to the azimuthally-averaged 21cm power spectrum, $\Delta T_b(k)^2 \equiv k^3 P_{21}(k)/ (2 \pi^2)$, when inhomogeneous X-ray heating dominates the fluctuations.  The lines correspond to the same star formation model as in Figure \ref{fig:Pk21}, with $\alpha =4$ for the minihalo cases.  Here we assume $x_\alpha \gg 1$, and that X-rays are responsible for raising the temperature to $\Tk = (T_{\rm ad} + \TCMB)/2$.  The fiducial choices maximize the amplitude of the component of the signal from X-ray heating.  The sensitivity curves for an MWA-like instrument are the same as described in Figure \ref{fig:Pkheating}.  \label{fig:Xray} \label{fig:PkX}}
\end{figure}

We argued in Section \ref{ss:temporal} that Ly$\alpha$ pumping with $\bar x_\alpha \sim 1$ is likely to have occurred before X-rays significantly heated the IGM, but, admittedly, there is much uncertainty in the high-redshift production of X-rays.  However, the fluctuations from X-ray heating are less damped than those from the Ly$\alpha$ background because of the shorter mean free path of the X-ray photons that are responsible for such heating.  The amount of damping is small on all scales we plot for the spectral index of the background that is assumed ($-1$ in intensity per unit frequency), and we found that this result was generic over a range of spectral indices.  

Figure \ref{fig:Xray} shows predictions for $P_{21}$ when the fluctuations are primarily driven by $z\sim 20$ X-ray backgrounds.  For simplicity, we assume $x_\alpha \gg 1$ so that fluctuations in $x_\alpha$ are zero and $\Tk = (T_{\rm ad} + \TCMB)/2$.  The amplitude of our curves can be rescaled to other $x_\alpha$ and $\Tk$.  Values of $x_\alpha \gg 1$ would imply large Lyman-Werner backgrounds that would suppress star formation in the environments we consider.  However, note that changing $x_\alpha = 2$ to $x_\alpha =\infty$ has little impact on our predictions.  
In addition, we assumed that the critical velocity that can form stars is determined by equation~(\ref{eqn:alpha}) with $\alpha =4$ for the minihalo curves in this figure.  If we instead used $\alpha =6$ (which is also consistent with the modulation observed in our simulations), this results in a factor of $\approx 2$ increase in the normalization of these curves.

The fluctuations are larger on observable scales in the case in which X-ray heating drives the fluctuations compared to the case where the fluctuations originate from the ultraviolet background.  The $P_{v^2}$ contribution in both of the minihalo models in Figure \ref{fig:Xray} may be detectable with an MWA-like instrument.  However, Figure \ref{fig:Xray} illustrates that only in the optimistic model with $V_{\rm cool, 0} =3.7~$km~s$^{-1}$ is the $P_{v^2}$ contribution comparable to the $P_\delta$ contribution and then only at $k < 0.1~\cMpc^{-1}$. Thus, even in this optimistic case in which X--rays dominate the fluctuations, the signal is not dramatically enhanced. \\

Thus, the contribution of $P_{v^2}$ to the 21cm signal is largest when X-rays from minihalos reheated the Universe.  However, in models in which atomic cooling halos dominate the $z\sim 20$ X-ray background, the contribution to the 21cm power spectrum that is sourced by $v_\bc$ is likely to be insignificant.  

\section{Conclusions}

This study presented semi-analytic calculations, which were calibrated with large cosmological simulations, aimed at understanding the early Universe and its 21cm signatures.  We focused on $z\sim 20$, an epoch when fluctuations in the diffuse 21cm background are not necessarily more difficult to detect than these fluctuations from any other cosmic era (e.g., \citealt{pritchard08}).  A detection of this signal would provide a window into when the first stars formed and into a time before the IGM had been reheated by astrophysical sources.  


We focused on the question of whether the dark matter--baryon supersonic differential velocity, $v_\bc$ \citep{tseliakhovich10}, is likely to enhance the $z\sim 20$ 21cm signal.   In \citet{paperI}, we showed that typical values for $v_\bc$ have a dramatic impact on the morphology of structures on scales of $10-100~$comoving kpc.  Whether $v_\bc$ impacted the 21cm signal in an observable manner boils down to whether $v_\bc$ also affected the $\sim 10\,$comoving~Mpc modes to which interferometric 21cm efforts are anticipated to be sensitive.  In fact, we showed that if just $\sim 1\%$ fractional fluctuations in the gas temperature were correlated with the large-scale $v_\bc$ flows, this coupling could lead to a significant new component to the signal.   We investigated three sources of $v_\bc$--coupled 21cm fluctuations relevant to the epoch after the first stars turned on but prior to reionization: (1) heating from structure formation and its associated shocks, (2) ultraviolet pumping of the 21cm line from the first stars' emissions,  and (3) heating by the X-ray background produced by the first supernovae, X-ray binaries, and miniquasars.\footnote{We did not consider the impact of $v_\bc$ on the signal from the reionization epoch as the consensus is that this epoch was driven by star formation in the atomic cooling halos that are too massive to be impacted by $v_\bc$ (but see \citealt{dalal10} and \citealt{bittner11}).  $v_\bc$ would also impact the clumpiness of gas and, thus, the number of recombinations during reionization (Joanne Cohn, private conversation).  This suppression would delay reionization in regions with smaller $v_\bc$.  However, any X-ray preheating prior to reionization would act to eliminate the clumpiness on scales impacted by $v_\bc$, and it is thought that $\sim10^7~\Msun$ minihalos -- which are only moderately impacted by $v_\bc$ -- are likely to dominate the number of absorptions from minihalos in all reionization scenarios \citep{iliev05, mcquinn07}.  Nevertheless, an investigation of this coupling mechanism would be interesting.}  

We found that the first of these sources, $v_\bc$'s impact on structure formation and shocking, did not contribute a significant level of large-scale 21cm fluctuations.  In particular, we found that simulations both with and without $v_\bc$ did not yield large enough differences in the average gas temperatures to result in fluctuations that were comparable to the fluctuations in the 21cm signal from density fluctuations.  We estimated that shocking contributes $\lesssim 10\%$ of the power at all wavenumbers.   This finding held true despite the fact that most of the overdense gas was decelerated into the potential wells of the dark matter via dynamical friction in our simulations, losing its relative velocity by $z\sim 20$.  This process leads to supersonic wakes and shocking throughout the cosmic volume, but not to significant heating. 

The other mechanism by which $v_\bc$ could imprint new fluctuations in the 21cm background involves the spatial modulation by $v_\bc$ of the formation of the first stars and their associated radiation backgrounds.  We investigated with a suite of cosmological simulations whether it is plausible that $v_\bc$ modulates early star formation.
At $z=20$, we found $\approx 3$ times more mass in gas that could cool in $1.0$ (or $0.1$) Hubble times and form stars in our simulations with $\calM_\bc =0$ than in those with $\calM_\bc = 1.9$.   Surprisingly, even the amount of dense gas in some of our most massive simulated halos  ($10^6-10^7~\Msun$) could be suppressed at the order--unity level for some halos in the simulations with $\calM_\bc = 1.9$.   We provided simplistic analytic estimates that reproduced the average amount of suppression in star-forming gas that was found in the simulations.  These estimates revealed why the impact is significant on halos with circular velocities of $\approx 10 \,v_\bc$, in contrast to previous estimates.  In detail, the amount of suppression in the simulations varied significantly from halo to halo at fixed halo mass and depended on, for example, the orientation of filamentary accretion flows with respect to the baryonic wind (see Paper I). 

The fluctuations in the 21cm background from inhomogeneous ultraviolet pumping are likely to be present at higher redshifts than fluctuations from other stellar-mediated radiation backgrounds.   
Unfortunately, the long mean free path of ultraviolet photons to redshift into the Ly$\alpha$ resonance dampens the  fluctuations from $v_{\bc}$ on observable scales.  We found that the $v_\bc$--dependence of ultraviolet pumping was a significant contribution to the 21cm anisotropy at $k\sim 0.01~$Mpc$^{-1}$, although not as large as in the models of \citet{dalal10} in which the impact of $v_\bc(z)$ on the star formation rate was rather extreme.  However, we found that $v_\bc$ could not source a significant component of the anisotropy on the largest observable scales ($k\gtrsim 0.1~$Mpc$^{-1}$) at $z=20$.  At $z=30$ (a redshift from which the diffuse 21cm background would be more difficult to observe), we found that the $v_\bc$ contribution at $k\sim 0.1~$Mpc$^{-1}$ could be more significant.

We concluded that $v_\bc$ is most likely to leave an ${\cal O}(1)$ imprint on the 21cm signal if X-ray heating from star-forming minihalos drives fluctuations in the intergalactic gas temperature.   Such X-ray production at high redshifts is very uncertain, and it is by no means guaranteed that X-ray reheating was sourced primarily by minihalos rather than by more massive halos that could have cooled atomically.  In fact, most previous models for the 21cm signal assumed the latter (e.g. \citealt{gnedin04, furlanetto06}).  X-ray reheating by minihalos would be more likely if minihalos produced more X-rays per unit star formation rate at $z=20$ than $z=0$ galaxies (such that X-ray reheating occurred before the Lyman-Werner background quenched star formation in minihalos).  Previous studies have made arguments for why this could have been the case \citep{oh01, mirabel11}.
In the cases we considered in which $> 3\times 10^5~\Msun$ and $> 3\times10^6~\Msun$ halos formed stars in proportion to their mass, we found that the 21cm anisotropy from $v_\bc$'s modulation of the X-ray background could be comparable to the 21cm signal sourced by density fluctuations at $z\sim 20$.  Most of this anisotropy was sourced by the impact of $v_\bc$ on the gas accreted by these halos and not from the suppression of the halo mass function (the effect considered in \citealt{tseliakhovich10}).  The acoustic oscillations in this $v_\bc$--sourced anisotropy make this finding particularly interesting as it results in a more distinctive signature for 21cm observatories to target -- a signature which would indicate that the Universe was reheated by star-forming minihalos.

Finally, we also investigated whether structure formation--initiated shocks reheated the Universe in Appendix A, which would reduce the absolute brightness temperature of the 21cm signal from $z\sim 20$, times when this line is anticipated to appear in absorption.  The one previous study that investigated this signal with cosmological simulations found that such shocks would dramatically suppress the amount of absorption \citep{gnedin04}.  The recently-funded LEDA and recently-proposed DARE 21cm observatories aim to detect this 21cm absorption trough in the sky-averaged 21cm signal, and the sensitivity of these efforts scales with the depth of the absorption trough.  Fortunately, we found that shock heating only suppresses the 21cm absorption at a fractional level of $\lesssim 20\%$ at $z>10$. \\

 We would especially like to thank Dusan Keres and Mike Kuhlen for their help with GADGET and Enzo.  We thank Rennan Barkana, Gianni Bernardi, Joanne Cohn, Lincoln Greenhill, Smadar Naoz, and Martin White for useful discussions.  We thank Volker Springel for  GADGET3.  Computations described in this work were performed using the Enzo code, developed by the Laboratory for Computational Astrophysics at the University of California in San Diego (\url{http://lca.ucsd.edu}), and with the yt analysis software \citep{yt}. MM and RO are supported by the National Aeronautics and Space
Administration through Einstein Postdoctoral Fellowship Award Number
  PF9-00065 (MM) and PF0-110078 (RO) issued by the Chandra X-ray Observatory Center, which is
operated by the Smithsonian Astrophysical Observatory for and on
behalf of the National Aeronautics Space Administration under contract
NAS8-03060.   This research was supported in part by the National Science Foundation through TeraGrid resources provided by the San Diego Supercomputing Center (SDSC) \citep{teragrid} and through award number AST/1106059.

\bibliographystyle{apj}
\bibliography{dynamical_heating}

\appendix

\section{A. When does shock heating result in the intergalactic temperature departing from adiabatic evolution?}
\label{ap:shocks}

\begin{figure}
\begin{center}
\rotatebox{-90}{\epsfig{file=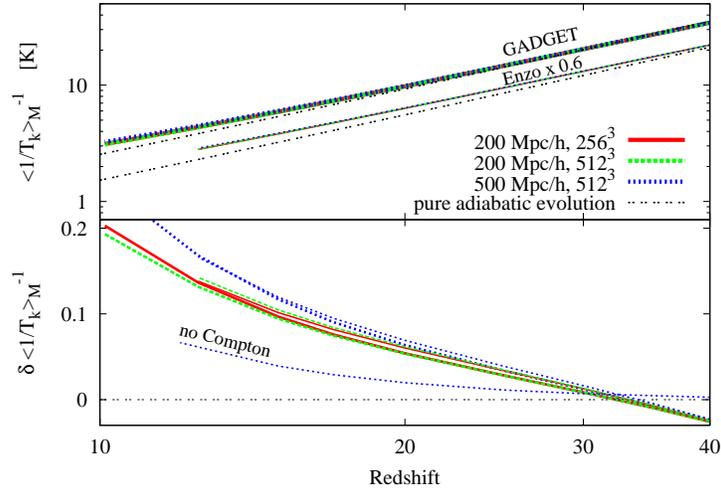, height=10cm}}
\end{center}
\caption{Evolution of the gas temperature in GADGET (thick curves) and Enzo (thin curves; multiplied by $0.6$ in the top panel) for simulations with the specified comoving box sizes and gas particle/grid dimensions.  The top panel shows  $\langle 1/\Tk \rangle_M^{-1}$, which is how the brightness temperature of the 21cm signal weights the temperature when $x_\alpha \gg 1$.  Also shown in this panel is the gas temperature for a homogeneous, expanding Universe normalized by eye (double dotted curves labelled ``pure adiabatic evolution'').  The bottom panel shows the fractional residuals of  $\langle 1/\Tk \rangle_M^{-1}$ with respect to the adiabatic evolution case.  Both GADGET and Enzo have almost identical residuals when compared at the same resolution and box size.  The curve labelled ``no Compton'' is from a simulation that did not include Compton heating but that was initialized with a temperature so that the simulated gas cooled to $10~$K at $z=20$.  \label{fig:Teff}}
\end{figure}

Here we quantify the importance of shock heating in determining the temperature history of the gas during the cosmic Dark Ages.  It is plausible that prior to X-ray reheating the Universe shock heated to significantly higher temperatures than if it had simply cooled adiabatically with the expansion of the Universe.  This heating would be analogous to what cosmological simulations find at $z\lesssim1$, where $\sim 50\%$ of the simulated mass shock heated to $\gtrsim 10^6~$K owing to $100~$km~s$^{-1}$ convergent flows \citep{cen99}.  However, at $z\sim 20$, a $0.3~$km~s$^{-1}$ flow was supersonic, so weak flows could have sourced shocks and heated the gas.

The middle and bottom panels in Figure \ref{fig:Teff} quantify how much heating occurs in our simulations.  The top panel shows $\langle \Tk^{-1} \rangle_M^{-1}$, where $\langle ... \rangle_M$ signifies a mass average, and the bottom panel shows the deviation of $\langle \Tk^{-1} \rangle_M^{-1}$ from adiabatic evolution.  Note that $\bar T^{21}_{b} \propto [1+ \TCMB \, \langle \Tk^{-1} \rangle_M]$ for $x_\alpha \, \TCMB/\Tk \gg 1$.  Thus, $\langle \Tk^{-1} \rangle_M^{-1}$ is the gas temperature-weighting most relevant to 21cm observations (after ultraviolet pumping becomes effective).   Both GADGET and Enzo have almost identical residuals when compared at the same resolution and box size.


Both codes predict $\approx 10-15\%$ deviations above adiabatic evolution by $z= 15$.  About half of this deviation owes to Compton heating rather than shocking.   (The ``no Compton'' curve does not include such heating.)  We have also run several adaptive simulations with Enzo that result in much higher resolution in overdense gas and find good convergence with the temperature evolution in these uni-grid Enzo calculations.  We conclude that structure formation shocks will not qualitatively change the 21cm signal, even in scenarios where the IGM is not impacted by astrophysics at $z\approx 10$.   

Two prior studies had looked at the importance of shock heating in setting the temperature of the early Universe.
\citet{furlanettoohbriggs} estimated the fraction of mass that shock heated above the CMB to be $f_{\rm sh} \sim \{0.1\%,~ 3\%,~ 25\%\}$ at $z = \{30,~ 20, ~10\}$ based on the amount of mass that was at turnaround in potential wells that were massive enough (i.e., had large enough characteristic velocities) to heat the IGM above the CMB temperature.  However, shocks could occur even prior to turnaround, as only $\sim 0.1-1~$km~s$^{-1}$ flows can be supersonic and shock.  For example, as shells cross in voids, this crossing will generate shocks in the gas (e.g., \citealt{bertschinger85}; although, shell crossing in spherical top-hat voids only occurs at a linear overdensity of $-2.7$).  In support of the possibility that shocks do not just occur at or after turnaround, the study of \citet{gnedin04} concluded using cosmological simulations that structure formation shocks are more prominent than the simple estimates in \citet{furlanettoohbriggs} suggest, creating order unity differences even at the highest redshift \citet{gnedin04} considered, $z=17$.  (Although, the simulations in \citealt{gnedin04} did not resolve the Jeans' scale.)  The \citet{furlanettoohbriggs} semi-analytic estimate is more consistent with the heating seen in our simulations.


\section{B. sensitivity to 21cm signal}
\label{ap:sensitivity}

The principle difficulty with detecting high-redshift 21cm radiation is that the sky is much brighter than the 21cm signal owing to foreground emission, the dominant foreground being synchrotron from the Galaxy.  Foreground emission not only must be subtracted off to isolate the 21cm signal, but also is likely to be the dominant source of statistical noise in a measurement of $P_{21}$.  The anticipated spectral smoothness should allow all significant extraterrestrial foregrounds to be separated from the 21cm signal.  However, since the brightness temperature of the sky scales as $T_{\rm sky} \approx 240 {\rm \, K \,}\left( \nu/150 ~\MHz \right)^{-2.55}$ \citep{rogers08}, it is thought that detecting the 21cm signal will be an even more difficult task as the targeted redshift increases (the observed frequency decreases; e.g., \citealt{mcquinn06}).  The challenge with statistically detecting the 21cm power spectrum at wavenumber $\bfk$ scales proportionally to $T_{\rm sky}^2/ \Delta \tilde T_{b, 21}^2(\bfk)$. 

To quantify the detectability of the signal in our models, we consider two hypothetical interferometer designs in this paper.  Our first design is motivated by the design of the MWA instrument, which is targeting the 21cm signal from $z\sim 10$.  
Specifically, for $z=20$ we assume an instrument that consists of a $40$ meter core of closely packed antennae with filling fraction $0.5$ and with an $r^{-2}$ baseline distribution to larger radii.  
We assume each dipole contributes $\lambda^2/4$ in collecting area and that the dipoles have spacing $\lambda/2$.   In addition, we contract or dilate this $z=20$ array in proportion to the targeted wavelength when we consider the signal from $z=15$ or $z=30$ in order to optimize the instrument to the specified redshift. 

MWA's cost is $\sim 20~$million US dollars for 500 antennae.  The cost of our analogous lower frequency instrument would be comparable.  We also consider in the body of this paper a next generation array with $10$ times the number of antennae elements and, for $z=20$, a $160~$meter core prior to an $r^{-2}$ falloff, which we denote as ``MWA--like~$\times$~10''.   The sensitivity of these two hypothetical instruments to the signal is discussed in Section \ref{sec:detectability}.  See \citet{mcquinn06} for additional details regarding our sensitivity calculations. 




\end{document}